\documentclass{aa}  
\usepackage{xcolor}
\usepackage{graphicx}
\usepackage{hyperref}
\hypersetup{
    colorlinks=true,
    linkcolor=blue,
    urlcolor=blue,
    citecolor=blue,
}
\usepackage{subfiles}
\usepackage{amssymb}
\usepackage{txfonts}
\newcommand{\kms}{km\,s$^{-1}$}
\newcommand{\CIV}{\ion{C}{IV}}
\newcommand{\SiIV}{\hbox{Si$\,\rm \scriptstyle IV$}}
\newcommand{\SiII}{\hbox{Si$\,\rm \scriptstyle II$}}
\newcommand{\CII}{\hbox{C$\,\rm \scriptstyle II$}}
\newcommand{\MgII}{\hbox{Mg$\,\rm \scriptstyle II$}}
\newcommand{\MgI}{\hbox{Mg$\,\rm \scriptstyle I$}}
\newcommand{\CaII}{\hbox{Ca$\,\rm \scriptstyle II$}}
\newcommand{\AlIII}{\hbox{Al$\,\rm \scriptstyle III$}}
\newcommand{\AlII}{\hbox{Al$\,\rm \scriptstyle II$}}
\newcommand{\FeII}{\hbox{Fe$\,\rm \scriptstyle II$}}
\newcommand{\HI}{\hbox{H$\,\rm \scriptstyle I$}}
\newcommand{\NaI}{\hbox{Na$\,\rm \scriptstyle I$}}
\newcommand{\ZnII}{\hbox{Zn$\,\rm \scriptstyle II$}}
\newcommand{\CrII}{\hbox{Cr$\,\rm \scriptstyle II$}}
\newcommand{\Lya}{Ly$\alpha$}

\begin{document}

   \title{Detecting the IGM metal enrichment with the 2-point correlation function of the flux}

   \subtitle{Application to the UVES deep spectrum}

   \author{S. Di Stefano{\inst{1,}\inst{2}} 
        \and V. D'Odorico{\inst{2,}\inst{3}} \and G. Cupani{\inst{2,}\inst{3}} \and D. Milakovic{\inst{2,}\inst{3}} \and A. Trost{\inst{1,}\inst{2,}\inst{4}} \and S. Cristiani{\inst{2,}\inst{3,}\inst{4}} \and  M. Viel{\inst{5,}\inst{2,}\inst{3,}\inst{4}} \and R. F. Carswell{\inst{6}}
          }

   \institute{Dipartimento di Fisica, Sezione di Astronomia, Università di Trieste, Via Tiepolo 11, I-34143 Trieste, Italy\\
              \email{simona.distefano@phd.units.it}
         \and
             INAF, Osservatorio Astronomico di Trieste,
              Via Tiepolo 11, I-34143 Trieste, Italy
        \and
            IFPU – Institute for Fundamental Physics of the Universe, 
            via Beirut 2,  I-34151 Trieste, Italy
        \and
            INFN – National Institute for Nuclear Physics, 
            via Valerio 2, I-34127 Trieste, Italy
        \and
            SISSA - International School for Advanced Studies, 
            via Bonomea 265, I-34136 Trieste, Italy    
        \and 
            Institute of Astronomy, University of Cambridge, Madingley Road, Cambridge CB3 0HA, UK
            }

   \date{Received date / Accepted date}
 
  \abstract
   {The distribution and the abundance of metals in the intergalactic medium (IGM) have strong implications on galaxy formation and evolution models. The ionic transitions of heavy elements in quasar spectra can be used to probe both the mechanisms and the sources of chemical pollution. However, the need for high-resolution and high signal-to-noise ratio (S/N) spectra makes it challenging to characterize the process of IGM metal enrichment, the IGM absorbers being too weak for direct detection.}
   {The aim of this work is to investigate the IGM metallicity, focusing on the detection of the weak absorption lines.}
   {We exploited the cosmological tool of the two-point correlation function (TPCF) and applied it to the transmitted flux in the \CIV\  forest region of the ultra-high S/N UVES spectrum of the quasar HE0940-1050 ($z \sim 3$). We also "deabsorbed" the strongest circum-galactic medium (CGM) systems in order to reveal the underlying IGM signal. For each of our tests, we created a catalogue of 1000 mock spectra in which we shuffled the position of the absorption lines, to derive an estimate for the TPCF noise level.}
   {The TPCF shows a clear peak at the characteristic velocity separation of the \CIV\  doublet. However, when removing the CGM contribution (i.e. deabsorbing all metal lines and \CIV\  lines associated with $\log N_{\rm{HI}} > 14.0$), the peak is not significant at 1 $\sigma$ anymore, even though 7 weak \CIV\  systems are still detectable by eye.
   Even after including up to 135 additional weak mock \CIV\ systems ($\log N_{\rm{HI}} < 14.0$) to the spectrum, we are not able to detect a significant \CIV\ peak.
   Eventually, when we create a synthetic spectrum with gaussian distributed noise and same S/N as the complete spectrum, we remove the signal caused by the spectral intrinsic features and thus find a peak compatible with a metallicity of $-3.80 < \rm{[C/H]} < -3.50$.} 
   {We conclude that the TPCF method is not sensitive to the presence of the weakest systems in the real spectrum, despite the extremely high S/N and high resolution of the data. However, the results of this statistical technique would possibly change when combining more than one line of sight.}


   \maketitle


\section{Introduction}
\label{sect:1_introduction}

The presence and distribution of metals in the intergalactic medium (IGM) are strongly related to the mechanisms and processes of galaxy formation and evolution. On the one hand, the material from the IGM fuels the birth and growth of galaxies; on the other hand, chemically enriched gas is recycled back into the circum-galactic medium (CGM) and IGM by means of feedback processes, such as stellar winds, supernova (SN) explosions and winds from active galactic nuclei (AGN). This cycle is known as the \textit{baryon cycle}. The gas that is continuously exchanged with the IGM and CGM is also progressively enriched in metals as the stars synthesize heavy elements and eject them. Models predict that, in particular in the early Universe, metal enriched gas can be distributed over large cosmological volumes, polluting the IGM to metallicities $[Z/Z_\odot] \simeq -3$ \citep{madau_ferrara_rees_2001}.

A possible way to observationally investigate the IGM metal enrichment through cosmic time is to use spectroscopy to search for its signatures along the line of sight to background bright quasars (see reviews by \citealt{becker+15}, \citealt{peroux_howk_2020}). Quasar absorption spectroscopy allows us to shed light on the presence of metals in both high and low ionization state, characterizing the enrichment history of the IGM up to high redshift values. Among the different transitions that one can find in a high-redshift quasar spectrum, the \CIV\ doublet ($\lambda \lambda \ 1548, 1550$ \AA) is definitely one of the most commonly detected, not only because carbon is one of the most abundant chemical elements in the Universe, but also because this ionic transition falls outside the \HI\ Lyman-$\alpha$ (Ly$\alpha$, hereafter) forest, where it is therefore generally free from line blending and easily recognizable due to its doublet nature.

With the use of cosmological hydrodynamic simulations, it is possible to trace the spatial distribution of metals in the galactic gaseous environments. \cite{cen_chisari_2011} show how effective the galactic superwind feedback from star formation is in transporting metals into the IGM, reaching a distance of influence of $\leq 0.5$ Mpc from the galaxy. 
They find that \CIV\ absorbers at $z = 2.6$ and $\log{N_{\rm{CIV}}} = 12 - 13$ generally arise in regions of overdensity\footnote{The overdensity is defined as $(\delta+1)= \rho / \bar{\rho}$, where $\rho$ is the mass density and $\bar{\rho}$ the cosmic mean mass density.} $1 + \delta \simeq 10$. On the other hand, \citet{shen+12,shen+13} focus on the CGM of single massive galaxies at $z \sim 3$.
These authors show that the metal enriched interstellar and circumgalactic material can extend up to 200 kpc from the centre of the galaxy, with the fraction of metals decreasing with redshift in the cold $T < 3 \times 10^4$ K phase of the gas (from $70\%$ at $z=8$ to $50\%$ ad $z=3$), increasing in the hot $T > 3 \times 10^5$ K gas (from $15\%$ at $z=8$ to $40\%$ ad $z=3$) and remaining constant in the warm $3 \times 10^4 \rm{K} < T < 3 \times 10^5 \rm{K}$ phase ($\sim 10\%$).
They attribute the metal enrichment mostly to the galaxy itself (60 \%), but also to its satellite progenitors and nearby dwarfs. In addition, \cite{shen+13} generated synthetic spectra by drawing sightlines through the CGM at different values of galactocentric impact parameters and observed the absorption lines of Ly$\alpha$ ($\lambda\ 1216$ \AA), \SiII\ ($\lambda\ 1360$ \AA), \SiIV\ ($\lambda\ 1393$ \AA), \CII\ ($\lambda\ 1334$ \AA), and \CIV. According to their results, the covering factor of the absorbing material declines less rapidly with the distance from the galaxy for \Lya\ and \CIV\ compared to \CII, \SiII\ and \SiIV.  

Observationally, the low density gas outside galaxies is traced by the low column density \HI\ lines in the Ly$\alpha$ forest ($\log{N_{\rm{HI}}} < $ 13.5 - 14.0). To study its metal content, one can adopt a direct approach by detecting weak metal absorption lines associated with low column density \HI\ lines, or statistical approaches which combine the signal pixel-by-pixel to detect the metals associated with the IGM. In both cases, high signal-to-noise ratio (S/N), high-resolution spectra are necessary.

An example of high S/N and high-resolution data is given by \cite{ellison+2000}, who study the spectrum of quasar Q1422+231 to investigate the \CIV\ enrichment associated with low \HI\ column densities in the Ly$\alpha$ forest. They do not recover any \CIV\ absorption signal in this regime by stacking the regions of the spectrum where \CIV\ absorptions associated with \HI\ lines would be expected. They applied also the so-called pixel optical depth method, which determines the \HI\ optical depth pixel-by-pixel and correlates it with the \CIV\ pixel optical depth at the corresponding redshifts. With this method, they find that detected \CIV\ absorption lines in the spectrum of Q1422+231 are not enough to explain the observed correlation between optical depths: additional \CIV\ lines below the detection limit are needed to match it, supporting the hypothesis that metals are present also in the low density gas. 

Similarly, \cite{schaye+03} applied the pixel optical depth method to a sample of high resolution and good S/N spectra (including the one studied by \citealt{ellison+2000}) and, using hydrodynamical simulations, transformed the correlation between \CIV\ and \HI\ optical depths into a relation between carbon abundance and overdensity as a function of redshift. The relation is a log normal distribution with a mild dependence on redshift. The median metallicity, expressed as [C/H], decreases with the overdensity and has a log value of  $-3.47$ at an overdensity of $\sim3$ and redshift $z=3$. 

It took more than 15 years, after the work of \citet{ellison+2000}, to collect another exceptionally high S/N and high resolution quasar spectrum to carry out this kind of studies.
\citet[][D16 hereafter]{d'odorico+16} analysed the VLT-UVES spectrum of the quasar HE0940-1050 ($z \sim 3$) to investigate the presence of \CIV\ absorptions associated with \HI\ absorbers at different column densities. At $\log{N_{\rm{HI}}} \geq 14.8$, all \HI\ lines have an associated \CIV\ absorption; the considered column density corresponds indicatively to an overdensity of $\delta \simeq 11$ \citep[applying the formula in ][see D16 for details]{schaye01}, which is usually assumed as the transition between the IGM and the CGM. At $14.0 \leq \log{N_{\rm{HI}}} < 14.8$, only 43\% of \HI\ lines shows an associated \CIV\ absorption and the estimated median metallicity falls in the range $ - 3 \lesssim [Z/Z_\odot] \lesssim -2.5 $, in rough agreement with the predictions of \citet{schaye+03} for the corresponding overdensities. Eventually, at $\log{N_{\rm{HI}}} \leq 14.0$ (corresponding to $\delta \simeq 2$) the detection rate drops to 10\%, partly because the possible associated \CIV\ lines would fall below the column density detection threshold, $\log N_{\rm CIV} = 11.4$.

In contrast to the direct detection of discrete absorption systems, another statistical technique to measure the chemical enrichment of the IGM was proposed by \citet{hennawi+21}, who used the cosmological tool of the two-point correlation function (TPCF) applied to the transmitted flux (see also \citealt{karacayli+23}). The signatures of the doublet transitions present in the spectrum are peaks of the TPCF at the velocity separation of the corresponding doublet (e.g. $\sim500$ \kms\ for \CIV\ or $\sim750$ \kms\ for \MgII\ $\lambda\,\lambda\ 2796, 2803$ \AA). \citet{hennawi+21} devised also a technique to filter the CGM contribution to the metal enrichment by using the probability distribution function (PDF) of $(1 - F)$, where $F$ is the normalized transmitted flux (see their paper for details). 

In \cite{tie+22}, they applied the method of \cite{hennawi+21} to mock spectra with \CIV\ absorptions at $z\sim 4.5$ obtained from simulations with a simplified model of inhomogeneous metal enrichment. The authors show that, adopting a mock sample of 20 UVES-like spectra and after filtering out the CGM metal absorptions, the proposed technique can constrain the IGM metallicity, [C/H], with a precision of 0.2 dex.

In this paper, we used the UVES deep spectrum from D16, which is also part of the UVES Spectral Quasar Absorption Database (SQUAD) Data Release 1 \citep{murphy+19}, and for which very weak \CIV\ systems associated with the IGM were directly detected, to perform a first observational test of the statistical method presented in \cite{tie+22}, namely the TPCF of the transmitted spectral flux in the \CIV\ forest. The goal of this work is to test the sensitivity of this technique in identifying the cumulative signature of very weak absorbers tracing the metals in the IGM, which could be mainly below the detection threshold of the data.  

The paper is organised as follows. In Section \ref{sect:2_dataset}, we briefly describe the quasar spectrum, as the best candidate to carry out our study. In Section \ref{sect:3_data_analysis}, we present the steps of the analysis, from the identification of the absorption lines to the computation of the two-point correlation function. Furthermore, we present our technique of deabsorption to filter out the CGM absorptions and how we determine the uncertainties on the measured correlation function. In Section \ref{sect:4_results}, we present our results. Eventually, the final discussion and some conclusions are drawn in Section \ref{sect:5_discussion_conclusions}.

\section{Dataset}
\label{sect:2_dataset}

This study is based on the UVES spectrum of the quasar HE0940-1050 (2MASSi J0942534-110425, $z_{\rm{em}} = 3.0932$), already presented in the work by D16 - which includes data from different observing programs (166.A-0106, 079.B-0469, 185.A-074, 092.A-0170) at the ESO VLT, for a total of $\sim 64$ hours of observation on source - and in \cite{murphy+19} as part of the Spectral QUasar Absorption Database (SQUAD) Data Release 1 - which includes data also from observing programs 65.O-0474 and 092.A-0770, for a total integration time of $\sim$ 73 hours. In this work, we used the SQUAD spectrum for our analysis and adopted the \CIV\ fitting parameters from D16.
\cite{murphy+19} reduced the data with a custom-written code, $\rm \scriptstyle UVES\textunderscore HEADSORT$, based on the ESO Common Pipeline Library (details of the data reduction are reported in \citealt{murphy16}). Exposures were then combined using a code designed by \citealt{murphy16b}, $\rm \scriptstyle UVES\textunderscore POPLER$, to produce a final spectrum, where the continuum was first fitted automatically with an iterative polynomial fit on selected spectral "chunks" and eventually adjusted manually around problematic regions, as the Ly$\alpha$ forest. 

The final spectrum (dubbed \textit{deep} spectrum in D16) is characterized by a high resolution ($R \sim 45,000$) and an extremely high S/N ($\sim 320-500$ per resolution element in the \CIV\ forest) and, therefore, it is chosen as the best candidate to test the sensitivity of the TPCF technique.

\section{Data analysis}
\label{sect:3_data_analysis}

\subsection{Line identification and fit}

\begin{figure*}[h!]
\centering
\includegraphics[scale=0.6]{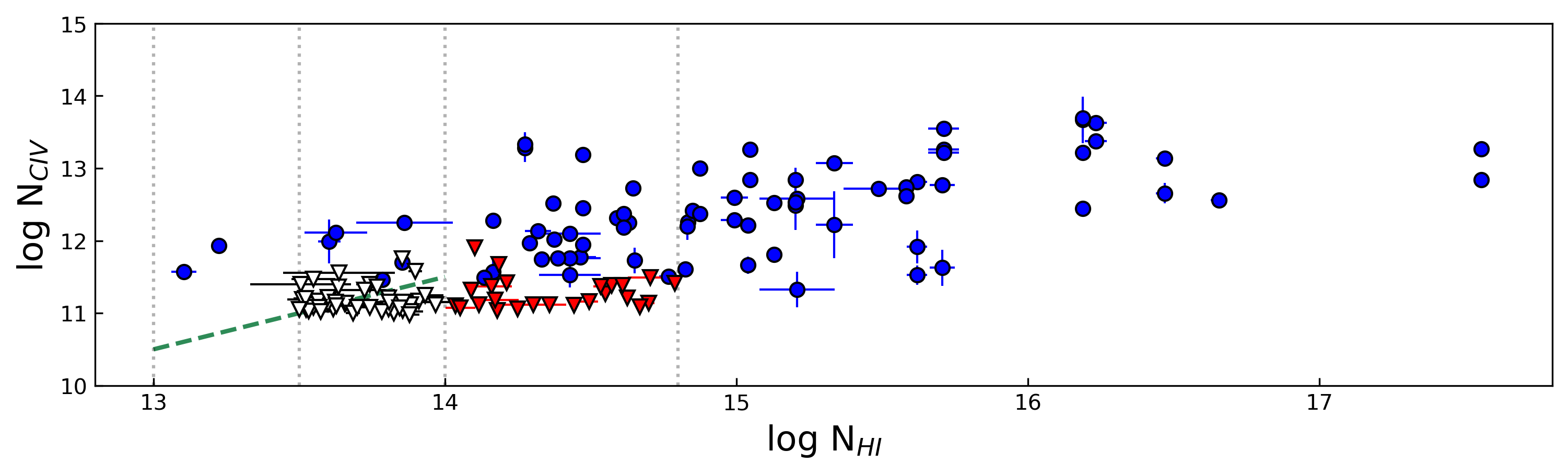}
\caption{\CIV\  column density, $\log N_{\rm{CIV}}$, versus the associated  column density of neutral hydrogen, $\log N_{\rm{HI}}$. The vertical dotted lines mark the values of $\log N_{\rm{HI}}= 13.0, 13.5, 14.0$ and $14.8$. Blue data points represent the detected \CIV\  systems identified by D16; white triangles indicate the upper limits on $\log N_{\rm{CIV}}$, which we used to derive the mock measurements (see Sect.~\ref{sect:test_uplim}). The red triangles represent the upper limits in the range between $\log N_{\rm{HI}}= 14.0$ and $\log N_{\rm{HI}}= 14.8$, not included in our tests. The green dashed line indicates the relation $\log N_{\rm{CIV}} = \log N_{\rm{HI}} - 2.5$ which is used, in the range $13 < \log N_{\rm{HI}} < 14$, to create the mock systems for one of our tests, see Sect.~\ref{sect:test_uplim}.}
\label{img:logNCIV_vs_logNHI}
\end{figure*}

The first step in the data analysis consists in identifying the spectral absorption lines, after careful visual inspection. In particular, we focused our analysis on the region of the spectrum outside the Ly$\alpha$ forest, redwards of the Ly$\alpha$ emission of the target. We analysed the \CIV\ systems in the wavelength window between 540 nm and 623 nm, corresponding to redshifts in the range $2.51 \lesssim z \lesssim 3.02$, where we excluded a proximity region of 5000 \kms\ from the quasar emission redshift. 

Line fitting was performed with the {\tt Astrocook} software, developed by \cite{cupani+20}, which provides the following parameters: the redshift of the absorbing cloud $z_{\rm{abs}}$, the logarithm of the column density $\log N$ and the Doppler parameter $b$, with their associated errors $\sigma_z$, $\sigma_{\rm{logN}}$ and $\sigma_b$. The software uses these parameters to create the line models, by fitting the absorption lines with Voigt profiles. In this work, we take the parameters for the \CIV\  systems from D16 (see Table \ref{tab:longtab_CIV}) and verify that they are consistent with the line fit performed with {\tt Astrocook}. In addition, we provide the identification and fitting parameters for all remaining metal lines.
In particular, we started by identifying the most common doublets and multiplets besides \CIV, such as: \MgII\ ($\lambda \lambda \ 2796, 2803$ \AA), \SiIV\ ($\lambda \lambda \ 1393, 1402$ \AA), \FeII\ ($\lambda \lambda\ 2344, 2382, 2586, 2600$ \AA), and \AlIII\ ($\lambda \lambda \ 1854, 1862$ \AA). Then, we looked for other transitions at the redshifts of the doublets due to e.g. \SiII, \CII\ ($\lambda\ 1334$ \AA),  \AlII\ ($\lambda\ 1670$ \AA), and \MgI\ ($\lambda\ 2852$ \AA). All detected lines are reported in Table~\ref{tab_metals}. 

D16 checked the presence of \CIV\ absorption associated with \Lya\ lines of decreasing column density, tracing the presence of metal enrichment in the IGM. For this reason, they fitted the \HI\ lines in the redshift range $2.49 \lesssim z \lesssim 3.02$, where at least the Ly$\alpha$ and the Ly$\beta$ transitions were available. Then, they found for each one of them the associated \CIV\ absorption (within a velocity window of 50 km s$^{-1}$) or, if the \CIV\ transition was not detected, they determined a $3\,\sigma$ upper limit on its column density at the redshift of the \HI\ line.  Column density upper limits were determined with equations 2 and 3 in D16, assuming the linear regime of the curve of growth and a Doppler parameter $b = 7$ \kms.  
The information regarding the fit of the \HI\ lines and of the corresponding \CIV\ lines are all reported in Table~\ref{tab:longtab_CIV}. In Fig. \ref{img:logNCIV_vs_logNHI} we report the \CIV\ column density versus the corresponding \HI\ column density, of both measurements (circles) and upper limits (white and red triangles). 

\subsection{Computation of the TPCF}

The two-point correlation function (TPCF) of the transmitted flux is computed, following \citet{tie+22}, as a function of the velocity separation between different pixels. 

Given two pixels with observed wavelengths $\lambda_1$ and $\lambda_2$, their wavelength separation can be converted into a velocity separation $\Delta v$ using the following relation, derived from the Doppler effect \citep[Eq. 3 in ][]{perrotta+16}, where $c$ is the speed of light:

\begin{equation}
    \Delta v = \left| \frac{\lambda_1^2 - \lambda_2^2}{\lambda_1^2 + \lambda_2^2} \right| c.
\end{equation}

Following Eq.~9 by \cite{tie+22}, the TPCF $\xi$ can be written as:

\begin{equation}
    \xi{(\Delta v)} = \frac{\sum_{i,j} w_i w_j \delta^{\rm{F}}_i \delta^{\rm{F}}_j}{\sum_{i,j} w_i w_j},
\end{equation}

\noindent 
where $w_i$ is the weight associated with pixel \textit{i}, computed as the inverse of the square of the error associated with the flux value of pixel \textit{i}. The TPCF is computed at the velocity separation $\Delta{v}$ of pixels $i$ and $j$. On the other hand, the flux overdensity $\delta^{\rm{F}}$ of pixel \textit{i} is defined as:

\begin{equation}
    \delta^{\rm{F}}_i = \frac{F_i - \bar{F}}{\bar{F}},
\end{equation}

\noindent
where $F_i$ is the normalized flux of pixel $i$ and $\bar{F}$ is the mean flux of the spectrum.

We computed the TPCF of the UVES deep spectrum of quasar HE0940-1050 considering the flux in the \CIV\ forest, corresponding to the wavelength range $540-623$ nm. This interval corresponds to a velocity range of $\sim 42 600$ \kms\ : since our velocity grid is set at much smaller scales (up to 3500 \kms), we are not missing substantial information when computing the TPCF. Having in mind the goal of isolating the signal due to the weak \CIV\ lines, we computed the TPCF also applying a process of "deabsorption" of the metal lines to the spectrum of HE0940-1050. Lines were deabsorbed subtracting the line model, obtained from the Voigt fitting in the context of {\tt Astrocook}, from the sum of spectral flux and the spectral continuum.\footnote{Another possibility to deabsorb the spectral lines would be to divide the flux by the line model, but we prefer to subtract it in order to avoid spikes when the flux and the model go to zero.} 
Uncertainties in the deabsorbed flux are the same as the original ones, as no error is associated with the line model.
By deabsorbing the lines, we remove their contribution to the final TPCF result. 

All the TPCFs described in this paper are computed on a velocity grid ranging from 0 \kms to 3500 \kms, with a bin size of 40 \kms.

\subsection{Computation of the uncertainties}
\label{sect:mock}

 Since we computed the TPCF for a single spectrum, we could not use the variance of the measurements \citep[as in ][]{tie+24} or bootstrapping techniques to estimate the error bars on the correlation function. 
 
 Therefore, we determined the errors of the TPCF values by creating a set of $N = 1000$ mock spectra for each version of the spectrum that we considered, and we computed the TPCF on these mock spectra, in order to obtain a distribution of the TPCF values for each velocity bin. Here we describe the procedure we used to create the mock spectra.

First, all absorption lines (\CIV\ and other metal lines) are deabsorbed: a totally deabsorbed spectrum is created.
Then, the absorption lines considered for the computation of the TPCF (see Section~\ref{sect:4_results}) are added to the deabsorbed spectrum with shuffled positions, meaning that the same $\log N$ and $b$ of the original systems are kept and only $z$ is changed. In particular, lines that originally belonged to a doublet are no longer inserted as a doublet (e.g. \CIV\ 1548 $\AA$ is considered as independent of \CIV\ 1550 $\AA$ and assigned to a different position in the spectrum).

In this way, we preserve the original flux but destroy all possible correlations between the systems, so that, when computing the TPCF on the set of mock spectra, the plot will show just some random fluctuations around the zero level (i.e. noise), which we use to give an estimate of the errors. It is still possible to observe some spikes due to random correlations that originate after the shuffling, but these spikes are limited in number and do not affect the final distribution in a relevant way, the number of mock spectra being sufficiently high.

The creation of mock spectra was also performed with {\tt Astrocook}, which allows creating mock Voigt profiles (based on a list of given parameters $\log N$, $b$ and $z$) and subsequently to superimpose them on the spectrum.

Once we computed the TPCF on all these samples of mock spectra, we took as error the region between the 16th and the 84th percentile of the distribution of the mock TPCFs in each velocity bin, corresponding to an uncertainty of $1\,\sigma$. The $3\,\sigma$ uncertainty was also computed from the distribution and considered in our analysis.

\section{Results}
\label{sect:4_results}

\subsection{TPCF of the complete spectrum}
\label{sect:TPCF_complete}

\begin{figure}[h!]
    \centering
    \includegraphics[scale=0.5]{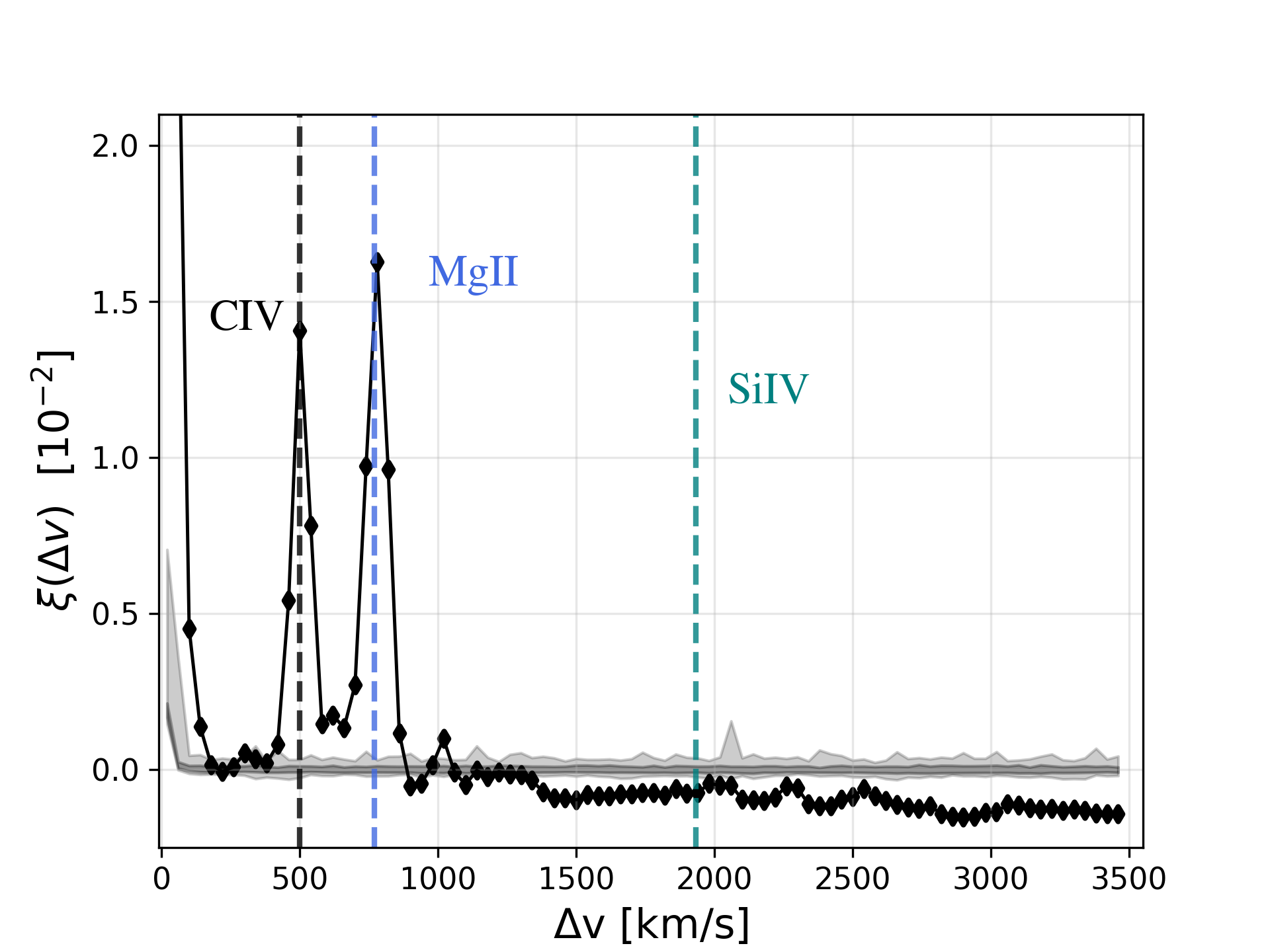}
    \caption{TPCF computed on the complete spectrum, in black. The dashed vertical lines are marking the velocity separation of the most common doublets: \CIV\  ($\sim$ 500 \kms), that is the one we are interested in, but also \MgII\ ($\sim$ 770 \kms) and \SiIV\ ($\sim$ 1933 \kms). The dark-gray and light-gray shaded regions indicate the uncertainties of $1\, \sigma$ and $3\, \sigma$ respectively.}
    \label{img:TPCF_original_spec}
\end{figure}

The TPCF computed with the complete spectrum  is shown in Fig. \ref{img:TPCF_original_spec}. The two peaks corresponding to the velocity separation of the \CIV\  doublet (at $\sim$ 500 \kms) and of the \MgII\ doublet (at $\sim$ 770 \kms) can be clearly seen in the plot, in addition to the peak at small velocity separations due to the structure of single lines or multi-component systems. Other peaks from other transitions (e.g. \SiIV, which is also marked in the plot) are not evident. We also observe a significant anti-correlation in the large-scale regime, which is explained by the presence of very strong absorption systems in our spectrum, characterized by large, negative $\delta^F$ values, anti-correlating with pixels at the continuum level.

We created a set of N = 1000 mock spectra by shuffling the position of both \CIV\  and other metal transitions. In Fig. \ref{img:TPCF_original_spec} we show the resulting $1\, \sigma$ uncertainty as a shaded region, with a lighter shaded region indicating an uncertainty of $3\, \sigma$. 

\subsection{TPCF of the deabsorbed spectra}

To isolate the contribution of the \CIV\ associated with the IGM, we compared two versions of the spectrum in which we performed a partial \CIV\  deabsorption.

In the first case, we deabsorbed all the metal lines except for the 31 \CIV\  lines associated with \HI\ lines with $\log N_{\rm{HI}}< 14.8$ (see Fig.~\ref{img:logNCIV_vs_logNHI}). Then, we computed the TPCF on the resulting spectrum. According to the relation by \citet[][see D16 for the details]{schaye01}, a column density of $\log N_{\rm{HI}}= 14.8$ corresponds at $z\simeq 2.8$ to an overdensity $(\delta+1) \simeq 11.8$, which is generally assumed as the interface between CGM and IGM. 
However, as shown in Section \ref{sect:4.3_flux_PDF}, by applying this cut to \CIV\  systems, the contribution from CGM absorbers is not entirely removed. 

In the second case, we deabsorbed all the metal lines and left only the weakest \CIV\  systems, namely those associated with \HI\ lines with $\log N_{\rm{HI}}< 14.0$, which are 7 systems in total. The TPCF was then computed also on this version of the spectrum. A column density of $\log N_{\rm{HI}} = 14.0$ at $z = 2.8$ is translated into an overdensity of $(\delta+1) \simeq 3.0$, typical of the intergalactic environment, as we confirm in Section \ref{sect:4.3_flux_PDF}.
In both cases, we determined the uncertainties on the TPCF as described in Section~\ref{sect:mock} and \ref{sect:TPCF_complete}, distributing over the totally deabsorbed spectra only the \CIV\ lines used for the computation of the respective TPCF.  

The two resulting TPCFs are shown in Fig.~\ref{img:TPCF_deabs_errors}. In the former case, the \CIV\ peak is visible and significant (i.e. out of the $3\,\sigma$ level), while in the latter case the TPCF does not show a significant peak at the velocity separation of the \CIV\ doublet. 

\subsection{TPCF of the deabsorbed spectra with mock weak systems}
\label{sect:test_uplim}

In Fig.~\ref{img:TPCF_deabs_errors} we show that, when we deabsorb all metal lines and leave only the \CIV\  associated with \HI\ systems with $\log N_{\rm{HI}}< 14.0$, there is no significant signal of IGM enrichment in the TPCF of the flux. Since there are 7 weak \CIV\ systems left in the spectrum after the deabsorption, this result suggests that the sensitivity of the TPCF technique, when applied to a real spectrum, is limited. 

To further test the sensitivity threshold of the method, we considered as detections the $3\, \sigma$ upper limits on the \CIV\ column density determined by D16 at the redshift of the \HI\ lines with $13.5 \le \log N_{\rm{HI}}< 14.0$ (see Table~\ref{tab:longtab_CIV} and the white triangles in Fig.~\ref{img:logNCIV_vs_logNHI}). Using {\tt Astrocook}, we created 46 mock \CIV\  Voigt profiles from the upper limits assuming a Doppler parameter $b=7$ \kms\ and we added them to the version of the spectrum with the weak \CIV\  systems associated with $\log N_{\rm{HI}}< 14.0$. 
The TPCF determined from this spectrum with the associated uncertainties is shown in Fig. \ref{img:TPCF_deabs_uCIV_mCIV} (left). 
Also in this case, the peak at 500 \kms\ is not significant.

As a final test, we adopted the procedure described in \citet{ellison+2000} to account for the metal enrichment associated also with Ly$\alpha$ lines with column densities of $13 <\log N_{\rm{HI}}< 13.5$ (89 lines in total), for which D16 did not compute the \CIV\ upper limits. 
We assumed the following relation between column densities: 

\begin{equation}
    \log N_{\rm{CIV}} = \log N_{\rm{HI}} - 2.5
\label{eq:logNHI_logNCIV_relation}
\end{equation}

\noindent
and applied it to all \Lya\ lines with $13.0 <\log N_{\rm{HI}}< 14.0$ (135 lines, see Fig.~\ref{img:logNCIV_vs_logNHI}), except those with a detected \CIV\ absorption.  Mock \CIV\ lines are generated assuming $b=7$ \kms\ and distributed on the spectrum with deabsorption at $\log N_{\rm{HI}}>14.0$. Finally, we computed the TPCF on the resulting spectrum and created the related set of N = 1000 mock spectra shuffling the lines as usual. The resulting TPCF and associated uncertainty region is shown in Fig. \ref{img:TPCF_deabs_uCIV_mCIV} (right). The peak is still within the shaded region, and therefore is not significant.

\subsection{TPCF of the synthetic spectra with weak systems}

\begin{figure*}[h!]
\centering
\includegraphics[scale=0.75]{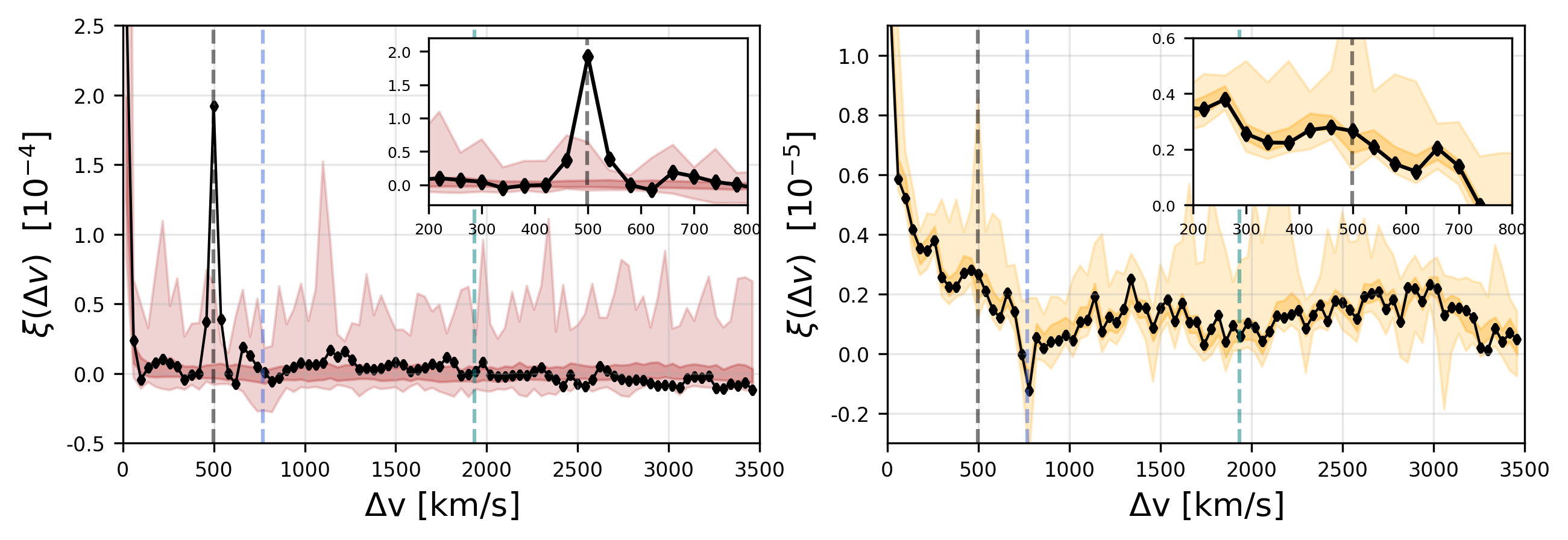}
\caption{TPCF for the spectrum where all metal lines have been deabsorbed except for \CIV\ systems  associated with $\log N_{\rm{HI}} < 14.8$ (\textit{left}) and $\log N_{\rm{HI}}< 14.0$ (\textit{right}) \Lya\ lines, plotted together with the $1\,\sigma$ and $3\,\sigma$ shaded regions obtained from the corresponding set of mock spectra.}
\label{img:TPCF_deabs_errors}
\end{figure*}

\begin{figure*}[h!]
\centering
\includegraphics[scale=0.75]{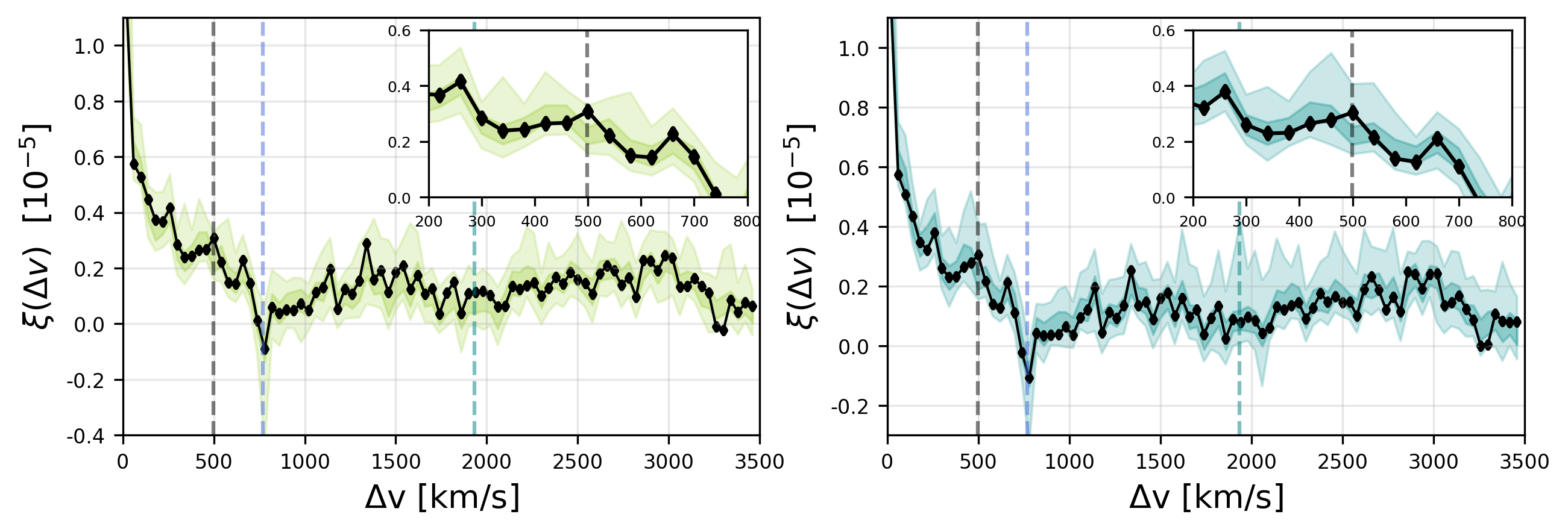}
\caption{\textit{Left:} TPCF computed on the spectrum in which we included the mock measurements derived from the upper limits in addition to the \CIV\  systems associated with $\log N_{\rm{HI}}< 14.0$. \textit{Right}: TPCF for the case in which we set the column density value of the mock \CIV\  measurements such that $\log N_{\rm{CIV}} = \log N_{\rm{HI}} - 2.5$ in the range $13 < \log N_{\rm{HI}} < 14$. The shaded coloured regions indicate the 1 $\sigma$ and 3 $\sigma$ regions obtained from the distributions of the TPCF values computed on the corresponding sets of mock spectra.}
\label{img:TPCF_deabs_uCIV_mCIV}
\end{figure*}

Another test we did to determine the sensitivity of the TPCF technique was to create synthetic spectra in which we inserted the extremely weak systems. This test is similar to the one described in the previous section, with the difference that we did not use the original spectrum: we created a synthetic spectrum which has the same S/N of the original totally deabsorbed one, but with gaussian distributed noise added to each pixel (this step was also performed with {\tt Astrocook}). Then, we added the \CIV\ systems associated with $\log N_{\rm{HI}}< 14.0$ in one case, and added the 135 mock systems (with column densities following Eq. \ref{eq:logNHI_logNCIV_relation}) in another case. In a sense, these two tests are analogous to Fig. \ref{img:TPCF_deabs_errors}, right, and Fig. \ref{img:TPCF_deabs_uCIV_mCIV}, right, but with the use of a synthetic spectrum, which allows us to remove the spurious signal that remains in the original flux, in order to highlight the strength of the \CIV\ peak that originates from the weak systems. This test is useful in order to understand how relevant the spectral features are in shaping the correlation function and see how the results change when we remove these features by mocking an "ideal deabsorption". On the other hand, we are also removing the possible signature of IGM enrichment present below the detection limit. 

\begin{figure*}[h!]
\centering
\includegraphics[scale=0.75]{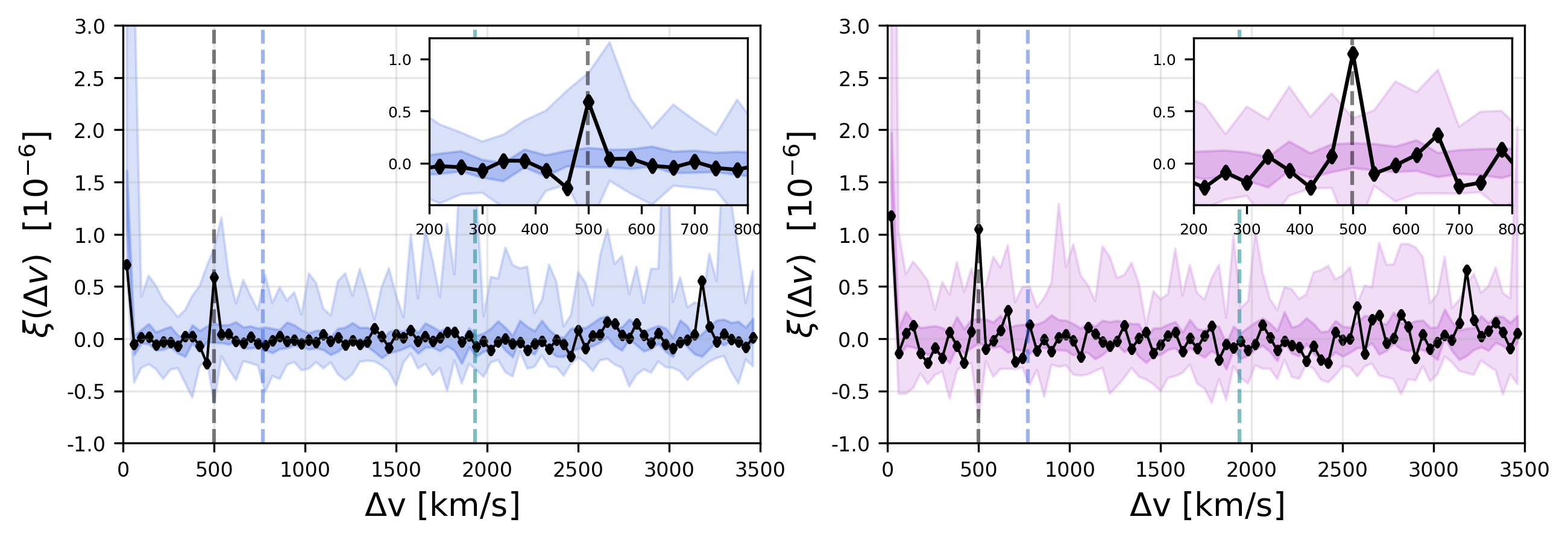}
\caption{\textit{Left:} TPCF computed on the synthetic spectrum (with same S/N as the original one) in which we included the \CIV\  systems associated with $\log N_{\rm{HI}}< 14.0$. \textit{Right}: TPCF for the case in which we also added the 135 mock measurements derived from Eq. \ref{eq:logNHI_logNCIV_relation} in addition to the weak systems. The shaded coloured regions indicate the 1 $\sigma$ and 3 $\sigma$ uncertainty regions obtained from the distributions of TPCF values computed on the corresponding sets of mock spectra.}
\label{img:TPCF_synth_uCIV_mCIV}
\end{figure*}

The resulting TPCFs are shown in Fig. \ref{img:TPCF_synth_uCIV_mCIV}. 
Results show a non-significant \CIV\ peak in the former case (Fig. \ref{img:TPCF_synth_uCIV_mCIV}, left), but a significant one (at a 3 $\sigma$ level) when we add the 135 mock measurements (Fig. \ref{img:TPCF_synth_uCIV_mCIV}, right). This suggests that the spectral features can considerably contaminate the TPCF signal and prevent the detection of the weak systems, this peak being not recorded by the previous test in Fig. \ref{img:TPCF_deabs_uCIV_mCIV}, right.
We also observe a significant peak at $\Delta v \sim 3180$ \kms, which we verified is due to the presence of the 4 weak \CIV\ doublets at $z \simeq 2.85796$, 2.89872, 2.940455 and 2.98251 with relative separations: $\Delta v \simeq 3150$, 3192 and 3182 \kms, respectively. 

\subsection{Flux probability distribution function}
\label{sect:4.3_flux_PDF}

In addition, we computed the flux decrement probability distribution function (PDF) in $\log_{10}$-unit, following the equation by \cite{tie+22}: 

\begin{equation}
    \text{frac}(a,b)=\int^b_a \frac{dP}{d \log_{10}(1-F)} d\log_{10}(1-F),
\end{equation}

\noindent
where $\text{frac}(a,b)$ is the fraction of pixels with $\log_{10}(1-F)$ values between $a$ and $b$. 

\citet{hennawi+21} and \citet{tie+22} use this distribution to identify the pixels in the \CIV\ forest associated with the CGM and mask them. Indeed, the CGM is expected to be associated with the strongest \CIV\ absorptions and therefore with the tail of the PDF at large $1-F$ values.

\begin{figure}[h!]
    \centering
    \includegraphics[scale=0.5]{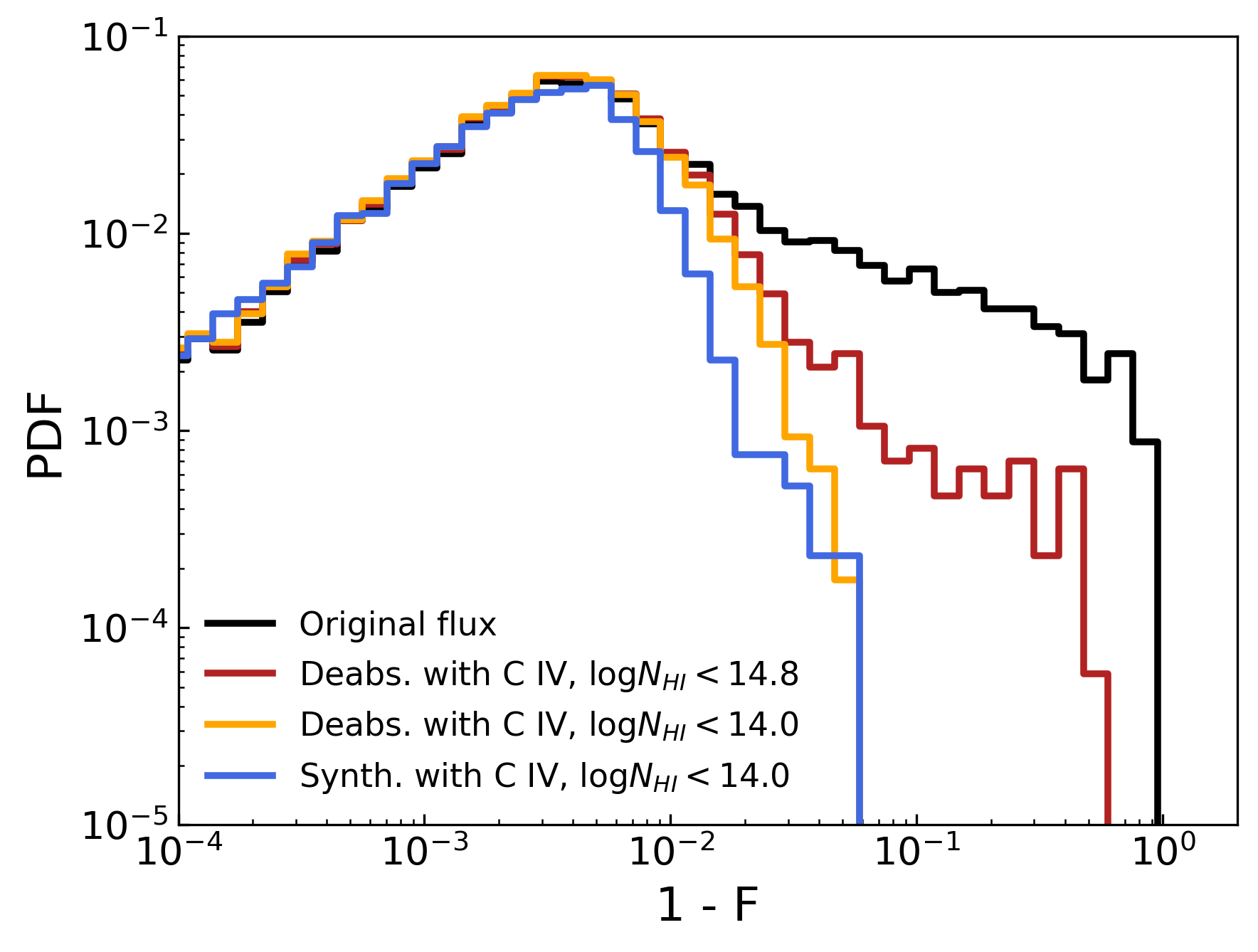}
    \caption{Flux PDF of four different versions of the spectrum: the original one (with no deabsorption, in black); the version with only \CIV\  systems associated with $\log N_{\rm{HI}}< 14.8$ (red); the one with only \CIV\  systems associated with $\log N_{\rm{HI}}< 14.0$ (orange); the synthetic spectrum with weak \CIV\  systems ($\log N_{\rm{HI}}< 14.0$) and same S/N as the original one (blue).}
    \label{img:PDF_spec}
\end{figure}

We checked the effects of our deabsorptions at different \HI\ column density thresholds on the PDF of the flux decrement. Figure \ref{img:PDF_spec} shows in black the PDF of the flux for the original spectrum with a clear tail extending to high values of $1-F$. We then computed the PDF for the spectrum with only \CIV\ associated with $\log N_{\rm{HI}} < 14.8$ (red line) and with $\log N_{\rm{HI}} < 14.0$ (orange line). In blue, we also show the PDF corresponding to the synthetic spectrum in Fig. \ref{img:TPCF_synth_uCIV_mCIV}, left.

We notice that when we deabsorb at $\log N_{\rm{HI}} = 14.8$ (red line) the CGM contamination is still present; on the other hand, the deabsorption at $\log N_{\rm{HI}} = 14.0$ (orange and blue lines) removes the CGM absorbers effectively.

We specify that the exact shape and normalization of the flux decrement PDF, and the value of $1-F$ at which there is the transition between IGM and CGM, may depend on the specific enrichment model adopted by \citet{tie+22}. However, since the shape and values of our PDF  are qualitatively consistent with the theoretical predictions, we consider it as a further empirical tool to identify the IGM-CGM transition.

\section{Summary and Discussion}
\label{sect:5_discussion_conclusions}

In this paper, we analysed the ultra-high S/N and high-resolution spectrum of the quasar HE0940-1050 ($z_{em} = 3.0932$), obtained with the VLT-UVES spectrograph, with the goal of detecting the IGM metal enrichment by computing the TPCF between the flux overdensities associated with the pixels, with a specific focus on the \CIV\  forest region at $2.51 \lesssim z \lesssim 3.02$ (corresponding to the wavelength interval 540 - 623 nm).

First, we identified and fitted with Voigt profiles the \CIV\ doublets and all the other metal absorption lines in the region of the spectrum redward of the Ly$\alpha$ emission of the quasar. Then, we computed the TPCF in the window 540 - 623 nm, for the following cases:  

\begin{description}
    \item{$i.$} original spectrum; 
    \item{$ii.$} spectrum deabsorbed of all metal lines except for the 31 \CIV\ lines associated with \Lya\ lines with $\log N_{\rm{HI}}< 14.8$;
    \item{$iii.$} spectrum deabsorbed of all metal lines except for the 7 weak \CIV\ lines associated with \Lya\ lines with $\log N_{\rm{HI}}< 14.0$; 
    \item{$iv.$} same as $iii.$, but with the addition of  46 mock measurements derived from the upper limits on $\log N_{\rm{CIV}}$ derived by D16; 
    \item{$v.$} same as $iii.$, but including also 135 mock measurements associated with the Ly-$\alpha$ lines with $13.0 < \log N_{\rm{HI}} < 14.0$ and such that $\log N_{\rm{CIV}} = \log N_{\rm{HI}} - 2.5$ and $b=7$ \kms;
    \item{$vi.$} same as $iii.$ but with a synthetic spectrum with the same S/N (gaussian distributed) as the totally deabsorbed complete spectrum;
    \item{$vii.$} same as $v.$ but using a synthetic spectrum with the same S/N (gaussian distributed) as the totally deabsorbed complete spectrum. 
\end{description}

For each of these cases, to estimate the errors associated with the TPCF values, we created a set of 1000 mock spectra by shuffling the position of the absorption lines which are left after the deabsorption procedure.

In case $i.$ (Fig. \ref{img:TPCF_original_spec}), the \CIV\  and \MgII\ peaks are clearly visible. We also observe the expected 
enhanced correlation on small separations, due to the velocity structure of the single absorption systems \citep{tie+24}, and its flattening at large velocity separation.  

In Fig. \ref{img:PDF_spec}, we show that applying the two different cuts (case $ii.$ and $iii.$) on the associated $\log N_{\rm{HI}}$, we gradually remove from the PDF the contribution from the strong CGM absorbers, which are responsible for the flatness of the function at large $1 - F$ values, until we keep only the weak IGM absorbers, which do not produce large flux decrements and therefore make the PDF drop at large $1 - F$ values \citep{hennawi+21, tie+22}. In particular, the PDF shows that deabsorbing only the \CIV\  systems associated with $\log N_{\rm{HI}}> 14.8$ (red line, corresponding to Fig. \ref{img:TPCF_deabs_errors}, left) is not sufficient to remove the CGM contribution, which still creates the flatness. We need to deabsorb more systems (orange line, corresponding to Fig. \ref{img:TPCF_deabs_errors}, right) in order to isolate the IGM signal. In other words, in case $ii.$, where the \CIV\  peak is significant at 3 $\sigma$, we are still affected by the systems associated with the CGM.
On the other hand, when we deabsorb even more \CIV\  systems (case $iii.$) and are left with the IGM contribution only, the \CIV\  peak is found to be consistent with the noise level (Fig. \ref{img:TPCF_deabs_errors}, right).

Figure \ref{img:TPCF_deabs_uCIV_mCIV} shows that the TPCF is not responsive to the presence of additional 135 weak mock \CIV\ systems with which we further enriched the spectrum, suggesting a lack of sensitivity related to systems associated with typical IGM column densities, $\log N_{\rm{HI}}< 14.0$.

The TPCFs in Fig. \ref{img:TPCF_deabs_errors} and Fig. \ref{img:TPCF_deabs_uCIV_mCIV} show a signal on scales smaller than $\sim 800$ \kms\ which we ascribe to the presence of residual flux after deabsorption and weak correlations in the noise structure. 
Indeed, when we use a synthetic spectrum in cases $vi.$ and $vii.$ (Fig. \ref{img:TPCF_synth_uCIV_mCIV}) the spurious signal disappears and the \CIV\ peaks due to the weak absorptions become visible. 
It is not clear whether applying the method to a larger sample would average out this spurious signal, which is always expected on the small scales. 

Finally, we can derive qualitative constraints on the IGM metal enrichment from the comparison of our results based on the synthetic spectrum with those obtained by \cite{tie+22} with noiseless spectra from simulations. Contrasting our Fig. \ref{img:TPCF_synth_uCIV_mCIV} with their Fig. 5, which shows the predictions they obtain for a resolution comparable to the UVES one (FWHM = 10 \kms), we can state the following:
\begin{description}
    \item{$\bullet$} assuming a minimum mass for the halos of $\log{M_{\rm{min}}} = 9.50 \ M_\odot$, and a radius of the enriched bubble (enrichment radius) of $R = 0.50$ cMpc, our result is consistent with a metallicity $\rm{[C/H]} \sim -3.80$;
    \item{$\bullet$} on the other hand, considering $R = 0.50$ cMpc and assuming $\rm{[C/H]} = -3.50$, we can estimate the minimum mass of the halo to be between $\log{M_{\rm{min}}} = 10$ and  $11 \ M_\odot$;
    \item{$\bullet$} finally, in the case of a minimum mass of $\log{M_{\rm{min}}} = 9.50 \ M_\odot$ and a metallicity $\rm{[C/H]} \sim -3.50$, our result indicates bubbles with a radius $R < 0.50$ cMpc. 
\end{description}

\noindent
The overall qualitative picture emerging from this comparison is that, assuming a metallicity $\rm{[C/H]} \sim -3.50$, the volume filling factor of the enriched regions predicted by the simple model by \citet{tie+22} to reproduce our TPCF is of the order of 5 per cent or less.  This is somewhat in contradiction with the result obtained by D16 (also adopting simple assumptions) where the volume filling factor of IGM gas enriched to a metallicity $\log Z/Z \gtrsim -3.0$  is derived to be $\sim10-13$ per cent.   
More detailed hydrodynamical simulations of the early chemical enrichment should be used to compare with observations and derive constraints on the metal enrichment of the IGM. 

In conclusion, with our study of a single line of sight at high resolution and very high S/N we show that, although very weak \CIV\ systems ($\log N_{\rm CIV}  \sim 11.5$) are present and detected by eye, their presence is not recorded by the TPCF technique. However, systems with column densities lower than $\log N_{\rm CIV}  \sim 11.7$ were not successfully recorded either with the stacking method in \cite{ellison+2000}, confirming the fact that the search for weak systems is a challenging task to pursue with even a very high S/N spectrum. 
We expect the results to improve when applying the same method to a larger sample of bright quasar spectra. However, residual correlations due to imperfections in the data reduction and analysis could affect the region of the \CIV\ peak and wash out the result even when multiple lines of sight are considered. We plan to extend the same procedure applied in this work to a larger sample of quasar spectra in a forthcoming paper.


\begin{acknowledgements}
     We are grateful to the anonymous referee for their comments, which have been useful to improve the quality of the paper. We would like to thank Louise Welsh and Vid Irsic for helpful discussions. This paper is based on observations collected at the European Southern Observatory Very Large Telescope, Cerro Paranal, Chile – Programs 65.O-0474, 166.A-0106, 079.B-0469, 185.A-0745, 092.A-0170 and 092.A-0770.

\end{acknowledgements}

\bibliographystyle{aa.bst}
\bibliography{0.bibliog}

\begin{thebibliography}{19}
\expandafter\ifx\csname natexlab\endcsname\relax\def\natexlab#1{#1}\fi

\bibitem[{{Becker} {et~al.}(2015){Becker}, {Bolton}, \& {Lidz}}]{becker+15}
{Becker}, G.~D., {Bolton}, J.~S., \& {Lidz}, A. 2015, \pasa, 32, e045

\bibitem[{{Cen} \& {Chisari}(2011)}]{cen_chisari_2011}
{Cen}, R. \& {Chisari}, N.~E. 2011, \apj, 731, 11

\bibitem[{{Cupani} {et~al.}(2020){Cupani}, {D'Odorico}, {Cristiani}, {Russo}, {Calderone}, \& {Taffoni}}]{cupani+20}
{Cupani}, G., {D'Odorico}, V., {Cristiani}, S., {et~al.} 2020, in Society of Photo-Optical Instrumentation Engineers (SPIE) Conference Series, Vol. 11452, Software and Cyberinfrastructure for Astronomy VI, ed. J.~C. {Guzman} \& J.~{Ibsen}, 114521U

\bibitem[{{D'Odorico} {et~al.}(2016){D'Odorico}, {Cristiani}, {Pomante}, {Carswell}, {Viel}, {Barai}, {Becker}, {Calura}, {Cupani}, {Fontanot}, {Haehnelt}, {Kim}, {Miralda-Escud{\'e}}, {Rorai}, {Tescari}, \& {Vanzella}}]{d'odorico+16}
{D'Odorico}, V., {Cristiani}, S., {Pomante}, E., {et~al.} 2016, \mnras, 463, 2690

\bibitem[{{Ellison} {et~al.}(2000){Ellison}, {Songaila}, {Schaye}, \& {Pettini}}]{ellison+2000}
{Ellison}, S.~L., {Songaila}, A., {Schaye}, J., \& {Pettini}, M. 2000, \aj, 120, 1175

\bibitem[{{Hennawi} {et~al.}(2021){Hennawi}, {Davies}, {Wang}, \& {O{\~n}orbe}}]{hennawi+21}
{Hennawi}, J.~F., {Davies}, F.~B., {Wang}, F., \& {O{\~n}orbe}, J. 2021, \mnras, 506, 2963

\bibitem[{{Kara{\c{c}}ayl{\i}} {et~al.}(2023){Kara{\c{c}}ayl{\i}}, {Martini}, {Weinberg}, {Ir{\v{s}}i{\v{c}}}, {Aguilar}, {Ahlen}, {Brooks}, {de la Macorra}, {Font-Ribera}, {Gontcho A Gontcho}, {Guy}, {Kisner}, {Miquel}, {Poppett}, {Ravoux}, {Schubnell}, {Tarl{\'e}}, {Weaver}, \& {Zhou}}]{karacayli+23}
{Kara{\c{c}}ayl{\i}}, N.~G., {Martini}, P., {Weinberg}, D.~H., {et~al.} 2023, \mnras, 522, 5980

\bibitem[{{Madau} {et~al.}(2001){Madau}, {Ferrara}, \& {Rees}}]{madau_ferrara_rees_2001}
{Madau}, P., {Ferrara}, A., \& {Rees}, M.~J. 2001, \apj, 555, 92

\bibitem[{{Murphy}(2016{\natexlab{a}})}]{murphy16}
{Murphy}, M. 2016{\natexlab{a}}, {UVES\_headsort: UVES\_headsort: VLT/UVES pipeline preparation}

\bibitem[{{Murphy}(2016{\natexlab{b}})}]{murphy16b}
{Murphy}, M. 2016{\natexlab{b}}, {UVES\_popler: Version 0.71 release}

\bibitem[{{Murphy} {et~al.}(2019){Murphy}, {Kacprzak}, {Savorgnan}, \& {Carswell}}]{murphy+19}
{Murphy}, M.~T., {Kacprzak}, G.~G., {Savorgnan}, G. A.~D., \& {Carswell}, R.~F. 2019, \mnras, 482, 3458

\bibitem[{{P{\'e}roux} \& {Howk}(2020)}]{peroux_howk_2020}
{P{\'e}roux}, C. \& {Howk}, J.~C. 2020, \araa, 58, 363

\bibitem[{{Perrotta} {et~al.}(2016){Perrotta}, {D'Odorico}, {Prochaska}, {Cristiani}, {Cupani}, {Ellison}, {L{\'o}pez}, {Becker}, {Berg}, {Christensen}, {Denney}, {Hamann}, {P{\^a}ris}, {Vestergaard}, \& {Worseck}}]{perrotta+16}
{Perrotta}, S., {D'Odorico}, V., {Prochaska}, J.~X., {et~al.} 2016, \mnras, 462, 3285

\bibitem[{{Schaye}(2001)}]{schaye01}
{Schaye}, J. 2001, \apj, 559, 507

\bibitem[{{Schaye} {et~al.}(2003){Schaye}, {Aguirre}, {Kim}, {Theuns}, {Rauch}, \& {Sargent}}]{schaye+03}
{Schaye}, J., {Aguirre}, A., {Kim}, T.-S., {et~al.} 2003, \apj, 596, 768

\bibitem[{{Shen} {et~al.}(2012){Shen}, {Madau}, {Aguirre}, {Guedes}, {Mayer}, \& {Wadsley}}]{shen+12}
{Shen}, S., {Madau}, P., {Aguirre}, A., {et~al.} 2012, \apj, 760, 50

\bibitem[{{Shen} {et~al.}(2013){Shen}, {Madau}, {Guedes}, {Mayer}, {Prochaska}, \& {Wadsley}}]{shen+13}
{Shen}, S., {Madau}, P., {Guedes}, J., {et~al.} 2013, \apj, 765, 89

\bibitem[{{Tie} {et~al.}(2022){Tie}, {Hennawi}, {Kakiichi}, \& {Bosman}}]{tie+22}
{Tie}, S.~S., {Hennawi}, J.~F., {Kakiichi}, K., \& {Bosman}, S. E.~I. 2022, \mnras, 515, 3656

\bibitem[{{Tie} {et~al.}(2024){Tie}, {Hennawi}, {Wang}, {Onorato}, {Yang}, {Ba{\~n}ados}, {Davies}, \& {O{\~n}orbe}}]{tie+24}
{Tie}, S.~S., {Hennawi}, J.~F., {Wang}, F., {et~al.} 2024, \mnras, 535, 223

\end{thebibliography}

\onecolumn

\begin{appendix}

\section{Tables}

\begin{longtable}{llllllllll}
\label{tab:longtab_CIV} \\
\caption{Parameters obtained by D16 from the fit of all \CIV\ lines. We also include the mock measurements derived from the upper limits, for which we set $b = 7$ \kms. The column density of mock measurements associated with $13 <\log N_{\rm{HI}}< 13.5$ is computed as $\log N_{\rm{CIV}} = \log N_{\rm{HI}} - 2.5$ (see Sect.~\ref{sect:test_uplim}).} \\
\hline \hline
$z_{\rm{HI}}$      & $\sigma_{z, \rm{HI}}$ & $\log N_{\rm{HI}}$ & $\sigma_{\rm{logN(HI)}}$ & $z$      & $\sigma_z$ & $\log N_{\rm{CIV}}$ & $\sigma_{\rm{logN(CIV)}}$ & $b$  & $\sigma_b$ \\
          &           &           &            &      &    &           &            & \kms & \kms    \\ \hline \\
\endfirsthead
\caption{continued.}\\
\hline \hline
$z_{\rm{HI}}$      & $\sigma_{z, \rm{HI}}$ & $\log N_{\rm{HI}}$ & $\sigma_{\rm{logN(HI)}}$ & $z$      & $\sigma_z$ & $\log N_{\rm{CIV}}$ & $\sigma_{\rm{logN(CIV)}}$ & $b$  & $\sigma_b$ \\
          &           &           &            &      &    &           &            & \kms & \kms    \\ \hline \\
\endhead
\\
\hline
\endfoot
2.492326 & 0.000005 & 13.486 & 0.056 & - & - & 10.986 & - & 7.0 & -  \\
2.493236 & 0.000003 & 13.060 & 0.029 & - & - & 10.560 & - & 7.0 & -\\
2.494527 & 0.000002 & 13.076 & 0.011 & - & - & 10.576 & - & 7.0 & -\\
2.496023 & 0.000018        & 13.55       & 0.08           & -      & -    & 11.47       & -            & 7.0 & -      \\
2.496298 & 0.000017        & 13.74       & 0.05           & -      & -    & 11.4       & -            & 7.0 & -      \\
2.500424 & 0.000005 & 13.415 & 0.045 & - & - & 10.915 & - & 7.0 & -\\
2.501682 & 0.000068 & 13.06 & 0.20 & - & - & 10.56 & - & 7.0 & -\\
2.502090 & 0.000006 & 13.299 & 0.074 & - & - & 10.799 & - & 7.0 & -\\
2.505407 & 0.000023 & 13.086 & 0.033 & - & - & 10.586 & - & 7.0 & -\\
2.506136 & 0.000002        & 13.910        & 0.004          & -      & -    & 11.17       & -            & 7.0 & -      \\
2.507588 & 0.000005        & 14.551        & 0.016          & -      & -    & 11.27       & -            & 7.0 & -      \\
2.512820 & 0.000003        & 13.635        & 0.005          & -      & -    & 11.36       & -            & 7.0 & -      \\
2.513387 & 0.000005 & 13.338 & 0.019 & - & - & 10.838 & - & 7.0 & -\\
2.515418 & 0.000009        & 14.495        & 0.031          & -      & -    & 11.16       & -            & 7.0 & -      \\
2.516268 & 0.000067        & 15.21       & 0.13           & 2.51609 & 0.00002   & 11.33       & 0.24           & 8.6 & 3.9    \\
          &           &           &            & 2.51623 & 0.00002   & 12.58       & 0.02           & 34.7 & 2.0    \\
2.516381 & 0.000008        & 14.37       & 0.02           & 2.516659 & 0.000001  & 12.51       & 0.02           & 9.0 & 0.3    \\
2.518442 & 0.000022 & 13.084 & 0.029 & - & - & 10.584 & - & 7.0 & -\\
2.524010 & 0.000002 & 13.099 & 0.004 & - & - & 10.599 & - & 7.0 & -\\
2.527235 & 0.000001        & 13.53       & 0.02           & -      & -    & 11.12       & -            & 7.0 & -      \\
2.529115 & 0.000008        & 14.77       & 0.03           & 2.52909 & 0.00001   & 11.51       & 0.08           & 6.2 & 2.0    \\
2.531654 & 0.000002        & 13.578        & 0.004          & -      & -    & 11.13       & -            & 7.0 & -      \\
2.534872 & 0.000017 & 13.458 & 0.076 & - & - & 10.958 & - & 7.0 & -\\
2.536117 & 0.000027        & 13.83       & 0.03           & -      & -    & 11.07       & -            & 7.0 & -      \\
2.541853 & 0.000020        & 13.60       & 0.14           & -      & -    & 11.19       & -            & 7.0 & -      \\
2.542630 & 0.000041        & 14.30       & 0.11           & -      & -    & 11.12       & -            & 7.0 & -      \\
2.543631 & 0.000002 & 13.282 & 0.009 & - & - & 10.782 & - & 7.0 & -\\
2.545325 & 0.000006 & 13.263 & 0.019 & - & - & 10.763 & - & 7.0 & -\\
2.551370 & 0.000004        & 14.699        & 0.020          & -      & -    & 11.14       & -            & 7.0 & -      \\
2.565277 & 0.000009 & 13.226 & 0.024 & - & - & 10.726 & - & 7.0 & -\\
2.566448 & 0.000001        & 14.290        & 0.003          & 2.566401 & 0.000005  & 11.97       & 0.03           & 6.0 & 0.7    \\
2.565655 & 0.000005 & 13.461 & 0.015 & - & - & 10.961 & - & 7.0 & -\\
2.567313 & 0.000003 & 13.276 & 0.007 & - & - & 10.776 & - & 7.0 & -\\
2.581817 & 0.000002 & 13.157 & 0.005 & - & - & 10.657 & - & 7.0 & -\\
2.592606 & 0.000010 & 13.186 & 0.026 & - & - & 10.686 & - & 7.0 & -\\
2.593969 & 0.000005        & 13.81       & 0.02           & -      & -    & 11.21       & -            & 7.0 & -      \\
2.594963 & 0.000008        & 14.625        & 0.008          & -      & -    & 11.21       & -            & 7.0 & -      \\
2.596428 & 0.000004        & 14.33       & 0.02           & 2.59644 & 0.00002   & 11.74       & 0.05           & 15.9 & 2.3    \\
2.596625 & 0.000065        & 13.93       & 0.06           & -      & -    & 11.25       & -            & 7.0 & -      \\
2.603414 & 0.000028 & 13.094 & 0.029 & - & - & 10.594 & - & 7.0 & -\\
2.604412 & 0.000008 & 13.456 & 0.012 & - & - & 10.956 & - & 7.0 & -\\
2.611654 & 0.000005        & 13.617        & 0.005          & -      & -    & 11.06       & -            & 7.0 & -      \\
2.612579 & 0.000003        & 14.179        & 0.005          & -      & -    & 11.04       & -            & 7.0 & -      \\
2.613643 & 0.000002        & 15.13       & 0.01           & 2.613624 & 0.000002  & 12.52       & 0.01           & 9.3 & 0.3    \\
          &           &           &            & 2.61389 & 0.00001   & 11.81       & 0.03           & 8.6 & 0.9    \\
2.616038 & 0.000016 & 13.033 & 0.039 & - & - & 10.533 & - & 7.0 & -\\
2.617323 & 0.000001        & 14.572        & 0.005          & -      & -    & 11.38       & -            & 7.0 & -      \\
2.622215 & 0.000001        & 14.789        & 0.007          & -      & -    & 11.41       & -            & 7.0 & -      \\
2.624459 & 0.000013 & 13.434 & 0.011 & - & - & 10.934 & - & 7.0 & -\\
2.626106 & 0.000058        & 13.73       & 0.10           & -      & -    & 11.32       & -            & 7.0 & -      \\
2.636121 & 0.000008 & 13.116 & 0.010 & - & - & 10.616 & - & 7.0 & -\\
2.643279 & 0.000015        & 15.20       & 0.02           & 2.64278 & 0.00002   & 12.49       & 0.06           & 15.7 & 1.8    \\
          &           &           &            & 2.643159 & 0.000004  & 12.84       & 0.16           & 9.4 & 1.1    \\
          &           &           &            & 2.64327 & 0.00012   & 12.53       & 0.38           & 17.1 & 5.3    \\
2.643685 & 0.000076        & 13.86       & 0.17           & 2.64371 & 0.00001   & 12.25       & 0.04           & 12.4 & 1.4    \\
2.644382 & 0.000005        & 14.609        & 0.011          & -      & -    & 11.38       & -            & 7.0 & -      \\
2.646742 & 0.000045 & 13.219 & 0.034 & - & - & 10.719 & - & 7.0 & -\\
2.647210 & 0.000005 & 13.219 & 0.029 & - & - & 10.719 & - & 7.0 & -\\
2.652885 & 0.000013 & 13.012 & 0.015 & - & - & 10.512 & - & 7.0 & -\\
2.654702 & 0.000007        & 14.534        & 0.025          & -      & -    & 11.37       & -            & 7.0 & -      \\
2.654839 & 0.000019        & 14.160        & 0.069          & -      & -    & 11.37       & -            & 7.0 & -      \\
2.655193 & 0.000045        & 13.51       & 0.17           & -      & -    & 11.4       & -            & 7.0 & -      \\
2.656691 & 0.000010        & 13.768        & 0.007          & -      & -    & 11.36       & -            & 7.0 & -      \\
2.657995 & 0.000001        & 15.05       & 0.01           & 2.657911 & 0.000002  & 13.26       & 0.01           & 14.3 & 0.2    \\
          &           &           &            & 2.657949 & 0.000001  & 12.84       & 0.02           & 4.9 & 0.3    \\
2.659215 & 0.000006        & 13.22       & 0.01           & 2.659252 & 0.000006  & 11.93       & 0.03           & 6.9 & 0.8    \\
2.660260 & 0.000002 & 13.285 & 0.004 & - & - & 10.785 & - & 7.0 & -\\
2.667487 & 0.000055        & 14.27       & 0.03           & 2.66744 & 0.00009   & 13.28       & 0.19           & 24.0 & 3.3    \\
          &           &           &            & 2.667634 & 0.000004  & 13.33       & 0.17           & 14.8 & 1.5    \\
2.667512 & 0.000001        & 16.47       & 0.03           & 2.667859 & 0.000005  & 12.66       & 0.14           & 4.2 & 0.8    \\
          &           &           &            & 2.667971 & 0.000009  & 13.14       & 0.05           & 8.5 & 0.6    \\
2.668987 & 0.000002        & 14.211        & 0.009          & -      & -    & 11.42       & -            & 7.0 & -      \\
2.670233 & 0.000027        & 14.705        & 0.078          & -      & -    & 11.49       & -            & 7.0 & -      \\
2.672484 & 0.000007 & 13.235 & 0.032 & - & - & 10.735 & - & 7.0 & -\\
2.673414 & 0.000002        & 14.185        & 0.010          & -      & -    & 11.67       & -            & 7.0 & -      \\
2.679609 & 0.000013 & 13.000 & 0.026 & - & - & 10.500 & - & 7.0 & -\\
2.684603 & 0.000004        & 13.524        & 0.006          & -      & -    & 11.05       & -            & 7.0 & -      \\
          &           & 13.548        & 0.003          & -      & -    & 11.08       & -            & 7.0 & -      \\
          &           & 13.856        & 0.003          & -      & -    & 11.04       & -            & 7.0 & -      \\
          &           & 13.660        & 0.004          & -      & -    & 11.13       & -            & 7.0 & -      \\
          &           & 13.599        & 0.003          & -      & -    & 11.13       & -            & 7.0 & -      \\
2.685215 & 0.000006 & 13.245 & 0.011 & - & - & 10.745 & - & 7.0 & -\\
2.687294 & 0.000004 & 13.429 & 0.032 & - & - & 10.929 & - & 7.0 & -\\
2.687445 & 0.000044 & 13.019 & 0.076 & - & - & 10.519 & - & 7.0 & -\\
2.705568 & 0.000001        & 14.036        & 0.003          & -      & -    & 11.10       & -            & 7.0 & -      \\
2.709389 & 0.000003 & 13.081 & 0.006 & - & - & 10.581 & - & 7.0 & -\\
2.711718 & 0.000001        & 14.834        & 0.004          & 2.71167 & 0.00002   & 11.61       & 0.06           & 10 & 2      \\
          &           & 13.825        & 0.006          & -      & -    & 11.0       & -            & 7.0 & -      \\ 
2.714507 & 0.000004 & 13.482 & 0.010 & - & - & 10.982 & - & 7.0 & -\\
2.726900 & 0.000005 & 13.428 & 0.008 & - & - & 10.928 & - & 7.0 & -\\
2.727577 & 0.000011 & 13.065 & 0.032 & - & - & 10.565 & - & 7.0 & -\\
2.739037 & 0.000002  & 13.51    &  0.01            & -         & -             & 11.19     & -                    & 7.0  & -       \\
2.739050 & 0.000005 & 13.483 & 0.029 & - & - & 10.983 & - & 7.0 & -\\
2.755819 & 0.000016 & 13.032 & 0.018 & - & - & 10.532 & - & 7.0 & -\\
2.756993 & 0.000002  & 13.882   &  0.004         & -        & -           & 11.12     & -                    & 7.0  & -       \\
2.758169 & 0.000002  & 13.660   &  0.004         & -         &  -          & 11.14     & -                    & 7.0  & -       \\
2.760091 & 0.000008 & 13.006 & 0.016 & - & - & 10.506 & - & 7.0 & -\\
2.761194 & 0.000002 & 13.498 & 0.005 & - & - & 10.998 & - & 7.0 & -\\
2.762458 & 0.000070 & 13.132 & 0.054 & - & - & 10.632 & - & 7.0 & -\\
2.762938 & 0.000003  & 13.56    &  0.03          & -         & -           & 11.17     & -                    & 7.0  & -       \\
2.769305 & 0.000004 & 13.102 & 0.010 & - & - & 10.602 & - & 7.0 & -\\
2.771169 & 0.000053  & 13.60    & 0.04           & 2.77126   & 0.00009    & 11.99     & 0.30                 & 12.9 & 4.9        \\
2.771364 & 0.000001  & 14.834   & 0.004          & 2.771410  & 0.000006   & 12.26     & 0.16                 & 6.8  & 1.1        \\
2.776551 & 0.000005 & 13.257 & 0.022 & - & - & 10.757 & - & 7.0 & -\\
2.777624 & 0.000005 & 13.363 & 0.015 & - & - & 10.863 & - & 7.0 & -\\
2.779332 & 0.000005 & 13.111 & 0.009 & - & - & 10.611 & - & 7.0 & -\\
2.785701 & 0.000027  & 13.86    &  0.21           & -          & -           & 11.15     & -                    & 7.0  & -       \\
2.785927 & 0.000021  & 14.47    & 0.05            & 2.78593   & 0.00001    & 11.77     & 0.04                 & 12.2 & 0.4        \\
2.788806 & 0.000002  & 13.532   &  0.006          & -          & -          & 11.03     & -                    & 7.0  & -       \\
2.790682 & 0.000005 & 13.227 & 0.009 & - & - & 10.727 & - & 7.0 & -\\
2.793037 & 0.000007 & 13.336 & 0.025 & - & - & 10.836 & - & 7.0 & -\\
2.795370 & 0.000022 & 13.249 & 0.095 & - & - & 10.749 & - & 7.0 & -\\
2.801124 & 0.000037 & 13.325 & 0.102 & - & - & 10.825 & - & 7.0 & -\\
2.810302 & 0.000011  & 13.90    &  0.02           & -          & -       & 11.58     & -                    & 7.0  & -       \\
2.810882 & 0.000002 & 14.849   & 0.003               & 2.810886  & 0.000004   & 12.41     & 0.02                 & 10.6 & 0.5        \\
2.812029 & 0.000009              & 14.102             & 0.004                    & -        & -          & 11.9               & -                    & 7.0  & -          \\
2.813041 & 0.000009 & 13.011 & 0.019 & - & - & 10.511 & - & 7.0 & -\\
2.822009 & 0.000014 & 13.60   &  0.04              & -        & -       & 11.22     & -                    & 7.0  & -       \\
2.822729 & 0.000011 & 14.38    & 0.03                & 2.823005  & 0.000008   & 12.02     & 0.03                 & 9.8  & 1.1        \\
2.823870 & 0.000024 & 15.62    & 0.04                & 2.82330   & 0.00002    & 11.52     & 0.13                 & 6.9  & 2.5        \\
         &          &          &                     & 2.82384   & 0.00002    & 12.81     & 0.04                 & 24.7 & 2.2        \\
         &          &          &                     & 2.82406   & 0.00001    & 11.92     & 0.23                 & 5.9  & 1.8        \\
2.824343 & 0.000017 & 15.34    & 0.06                & 2.82428   & 0.00001    & 13.07     & 0.07                 & 11.0 & 1.0        \\
         &          &          &                     & 2.82449   & 0.00010    & 12.22     & 0.46                 & 11.5 & 5.7        \\
2.825109 & 0.000007 & 14.87    & 0.01                & 2.825117  & 0.000006   & 13.00     & 0.01                 & 24.8 & 1.0        \\
         &          &          &                     & 2.825149  & 0.000003   & 12.37     & 0.04                 & 6.2  & 0.5        \\ 
2.825592 & 0.000045 & 15.040   & 0.007               & 2.825459  & 0.000007   & 11.66     & 0.12                 & 4.4  & 1.4        \\
         &          &          &                     & 2.825924  & 0.000006   & 12.22     & 0.08                 & 10.1 & 1.1        \\
2.826528 & 0.000002 & 14.473   & 0.006               & 2.82634   & 0.00006    & 12.45     & 0.09                 & 29.8 & 6.5        \\
         &          &          &                     & 2.8265569 & 0.0000005  & 13.188    & 0.006                & 6.49 & 0.08       \\
         &          &          &                     & 2.826765  & 0.000005   & 11.95     & 0.08                 & 5.1  & 0.9        \\
2.827334 & 0.000011 & 14.99    & 0.05                & 2.827370  & 0.000004   & 12.29     & 0.04                 & 7.4  & 0.6        \\
         &          &          &                     & 2.827716  & 0.000002   & 12.60     & 0.02                 & 5.5  & 0.3        \\
2.828221 & 0.00002  & 16.23    & 0.04                & 2.8280270 & 0.0000008  & 13.38     & 0.01                 & 11.3 & 0.1        \\
         &          &          &                     & 2.828065  & 0.000008   & 13.630    & 0.009                & 51.5 & 1.0        \\
2.828353 & 0.000072 & 15.49    & 0.12                & 2.828344  & 0.000002   & 12.72     & 0.02                 & 7.4  & 0.3        \\
2.828912 & 0.000012 & 15.71    & 0.04                & 2.828868  & 0.000004   & 12.77     & 0.02                 & 9.1  & 0.3        \\
         &          &          &                     & 2.828939  & 0.000008   & 11.63     & 0.25                 & 2.5  & 2.1        \\
2.83135  & 0.000028 & 13.64    & 0.19                & -         & -          & 11.56     & -                    & 7.0  & -       \\
2.831860 & 0.000043 & 13.17 & 0.16 & - & - & 10.67 & - & 7.0 & -\\
2.832145 & 0.000014 & 13.10 & 0.17 & - & - & 10.60 & - & 7.0 & -\\
2.832798 & 0.000006 & 15.58    & 0.01                & 2.832878  & 0.000007   & 12.74     & 0.02                 & 17.7 & 0.7        \\
         &          &          &                     & 2.833322  & 0.000006   & 12.62     & 0.02                 & 14.4 & 0.7        \\
2.834181 & 0.000024 & 16.19    & 0.02                & 2.833965  & 0.000002   & 12.44     & 0.07                 & 5.3  & 0.6        \\
         &           &          &                     & 2.833976  & 0.000003   & 13.22     & 0.01                 & 16.4 & 0.5        \\
         &          &          &                     & 2.834512  & 0.000060   & 13.67     & 0.32                 & 13.8 & 1.0        \\
         &          &          &                     & 2.834557  & 0.000008   & 13.69     & 0.26                 & 8.8  & 1.1        \\   
2.834970 & 0.000021 & 15.71    & 0.05                & 2.834810  & 0.000005   & 13.55     & 0.07                 & 8.0  & 0.7        \\
& &          &                     & 2.834995  & 0.000006   & 13.26     & 0.09                 & 7.9  & 0.7        \\
& &          &                     & 2.83511   & 0.00004    & 13.27     & 0.11                 & 17.9 & 2.1        \\
2.840136 & 0.000008 & 13.254 & 0.011 & - & - & 10.754 & - & 7.0 & -\\
2.844643 & 0.000013 & 13.52   &  0.04              & -  & -   & 11.2      & -                    & 7.0  & -       \\
2.845361 & 0.000041 & 13.12 & 0.15 & - & - & 10.62 & - & 7.0 & -\\
2.846060 & 0.000030 & 13.376 & 0.061 & - & - & 10.876 & - & 7.0 & -\\
2.849489 & 0.000006 & 13.305 & 0.015 & - & - & 10.805 & - & 7.0 & -\\
2.853444 & 0.000008 & 13.365 & 0.013 & - & - & 10.865 & - & 7.0 & -\\
2.857865 & 0.000001 & 13.853   &  0.002              & -  & -   & 11.75     & -                    & 7.0  & -       \\
2.857865 & 0.000001 & 13.853   & 0.002               & 2.85796   & 0.00002    & 11.70     & 0.07                 & 13.9 & 3.0        \\
2.860654 & 0.000017 & 14.83    & 0.01                & 2.86075   & 0.00003    & 12.20     & 0.19                 & 7.9  & 1.6        \\
2.860862 & 0.000001 & 16.66    & 0.03                & 2.86091   & 0.00001    & 12.56     & 0.08                 & 7.8  & 0.9        \\
2.862071 & 0.000013 & 13.344 & 0.096 & - & - & 10.844 & - & 7.0 & -\\
2.862118 & 0.000052 & 13.62   &  0.07              & -   & -    & 11.15     & -                    & 7.0  & -       \\
2.862518 & 0.000032 & 13.143 & 0.084 & - & - & 10.643 & - & 7.0 & -\\
2.863317 & 0.000002              & 14.117             & 0.004                    & -        & -          & 11.12               & -                    & 7.0  & -          \\
2.864039 & 0.000021 & 13.183 & 0.077 & - & - & 10.683 & - & 7.0 & -\\
2.865753 & 0.000008 & 13.50   &  0.01              & -  & -    & 11.05     & -                    & 7.0  & -       \\
2.866758 & 0.000002              & 14.443             & 0.002                    & -        & -          & 11.11               & -                    & 7.0  & -          \\
2.867669 & 0.000045 & 13.44 & 0.10 & - & - & 10.94 & - & 7.0 & -\\
2.871513 & 0.000002              & 14.359             & 0.005                    & -        & -          & 11.12               & -                    & 7.0  & -          \\
2.871996 & 0.000009 & 13.97   &  0.02              & -   & -   & 11.12     & -                    & 7.0  & -       \\
2.874950 & 0.000003 & 13.279 & 0.041 & - & - & 10.779 & - & 7.0 & -\\
2.877392 & 0.000026 & 13.24 & 0.11 & - & - & 10.74 & - & 7.0 & -\\
2.878524 & 0.000006 & 13.46 & 0.18 & - & - & 10.96 & - & 7.0 & -\\
2.882940 & 0.000028 & 13.423 & 0.040 & - & - & 10.923 & - & 7.0 & -\\
2.883511 & 0.000002 & 14.631   & 0.004               & 2.883528  & 0.000003   & 12.25     & 0.01                 & 11.1 & 0.4        \\
2.893671 & 0.000002              & 14.669             & 0.007                    & -        & -          & 11.09               & -                    & 7.0  & -          \\
2.894198 & 0.000051 & 13.377 & 0.067 & - & - & 10.877 & - & 7.0 & -\\
2.898551 & 0.000008 & 13.10    & 0.04                & 2.89872   & 0.00002    & 11.57     & 0.07                 & 12.2 & 2.5        \\
2.899046 &  0.000008 & 13.68   &  0.03              & -  & -    & 11.0      & -                    & 7.0  & -       \\
2.899272 & 0.000002 & 13.81   &  0.01              & -   & -   & 11.06     & -                    & 7.0  & -       \\
2.900734 & 0.000012 & 13.316 & 0.021 & - & - & 10.816 & - & 7.0 & -\\
2.906392 & 0.000016 & 13.70   &  0.05              &  - & -   & 11.08     & -                    & 7.0  & -       \\
2.915405 & 0.000003              & 14.090             & 0.011                    & -        & -          & 11.32               & -                    & 7.0  & -          \\
2.916606 & 0.000041 & 14.65    & 0.02                & 2.916686  & 0.000005   & 11.73     & 0.18                 & 3.2  & 2.5        \\
2.916960 & 0.000002 & 17.555   & 0.003               & 2.91688   & 0.00002    & 13.27     & 0.02                 & 24.7 & 0.6        \\
         &          &          &                     & 2.917109  & 0.000003   & 12.84     & 0.05                 & 10.4 & 0.5        \\
2.918044 & 0.000037 & 14.43    & 0.11                & 2.91757   & 0.00001    & 11.53     & 0.17                 & 4.3  & 3.1        \\
         &          &          &                     & 2.91794   & 0.00002    & 12.10     & 0.05                 & 15.9 & 1.9        \\
         &          &          &                     & 2.918045  & 0.000005   & 11.76     & 0.10                 & 3.4  & 1.2        \\
2.920227 & 0.000016 & 13.37 & 0.18 & - & - & 10.874 & - & 7.0 & -\\
2.921212 & 0.000011 & 13.184 & 0.064 & - & - & 10.684 & - & 7.0 & -\\
2.921868 & 0.000010 & 13.242 & 0.057 & - & - & 10.742 & - & 7.0 & -\\
2.925395 & 0.000003 & 13.182 & 0.025 & - & - & 10.682 & - & 7.0 & -\\
2.925565 & 0.000038 & 13.266 & 0.012 & - & - & 10.766 & - & 7.0 & -\\
2.926851 & 0.000004 & 13.784   &  0.009              & -  & -    & 11.02     & -                    & 7.0  & -       \\
2.928590 & 0.000005 & 13.190 & 0.006 & - & - & 10.690 & - & 7.0 & -\\
2.929947 & 0.000054 & 13.102 & 0.088 & - & - & 10.602 & - & 7.0 & -\\
2.930796 & 0.000012 & 14.59    & 0.03                & 2.93080   & 0.00001    & 12.31     & 0.06                 & 11.2 & 1.3        \\
2.931025 & 0.000008 & 14.32    & 0.05                & 2.93106   & 0.00002    & 12.13     & 0.09                 & 9.5  & 1.6        \\
2.935827 & 0.000006 & 13.229 & 0.008 & - & - & 10.729 & - & 7.0 & -\\
2.937219 & 0.000011 & 14.17    & 0.01                & 2.93711   & 0.00002    & 12.28     & 0.04                 & 23.9 & 2.2        \\
         &          &          &                     & 2.93719   & 0.00001    & 11.57     & 0.17                 & 5.8  & 2.3        \\
2.937722 & 0.000002 & 14.646   & 0.005               & 2.937751  & 0.000002   & 12.723    & 0.007                & 10.6 & 0.2        \\
         &          & 14.61    & 0.01                & 2.939641  & 0.000001   & 12.371    & 0.006                & 4.7  & 0.1        \\
         &          &          &                     & 2.940080  & 0.000006   & 12.18     & 0.02                 & 13.1 & 0.7        \\         
2.940125 & 0.000006 & 13.63    & 0.11                & 2.940455  & 0.000004   & 12.11     & 0.02                 & 8.8  & 0.5        \\
2.945143 & 0.000009 & 13.014 & 0.029 & - & - & 10.514 & - & 7.0 & -\\
2.948594 & 0.000018              & 14.053             & 0.051                    & -        & -          & 11.07               & -                    & 7.0  & -          \\
2.948886 & 0.000039 & 13.63   &  0.14              & -  & -    & 11.1      & -                    & 7.0  & -       \\
2.949032 & 0.000012 & 13.57   &  0.02              & -   & -    & 11.08     & -                    & 7.0  & -       \\
2.950417 & 0.000010 & 14.13    & 0.02                & 2.95046   & 0.00002    & 11.49     & 0.06                 & 9.0  & 1.7        \\
2.950797 & 0.000018 & 13.88   &  0.04              & -   & -   & 11.02     & -                    & 7.0  & -       \\
2.957502 & 0.000004 & 13.346 & 0.016 & - & - & 10.846 & - & 7.0 & -\\
2.963150  &  0.000010 & 13.74   &  0.01              & -   & -   & 11.08     & -                    & 7.0  & -       \\
2.963563 & 0.000048 & 13.06 & 0.20 & - & - & 10.56 & - & 7.0 & -\\
2.975361 & 0.000008 & 13.06 & 0.10 & - & - & 10.56 & - & 7.0 & -\\
2.977992 & 0.000015 & 13.81   &  0.20              & -  & -   & 11.15     & -                    & 7.0  & -       \\
2.979092 & 0.000021 & 13.151 & 0.076 & - & - & 10.651 & - & 7.0 & -\\
2.978125 & 0.000033              & 14.173             & 0.080                    & -        & -          & 11.18               & -                    & 7.0  & -          \\
2.981341 & 0.000004 & 13.108 & 0.010 & - & - & 10.608 & - & 7.0 & -\\
2.982029 & 0.000025 & 13.489 & 0.080 & - & - & 10.989 & - & 7.0 & -\\
2.982516 & 0.000003 & 13.785   & 0.004               & 2.98251   & 0.00002    & 11.46     & 0.06                 & 12.1 & 2.4        \\
2.983844 & 0.000040 & 13.083 & 0.066 & - & - & 10.583 & - & 7.0 & -\\
2.984655 & 0.000006 & 13.216 & 0.045 & - & - & 10.716 & - & 7.0 & -\\
2.986108 & 0.000001 & 13.845   &  0.001              & -   & -   & 11.07     & -                    & 7.0  & -       \\
3.021212 & 0.000001              & 14.250             & 0.004                    & -        & -          & 11.06               & -                    & 7.0  & -          \\
3.002621 & 0.000014 & 13.227 & 0.030 & - & - & 10.727 & - & 7.0 & -\\
3.003015 & 0.000006 & 13.451 & 0.018 & - & - & 10.951 & - & 7.0 & -\\
3.007371 & 0.000015 & 13.073 & 0.060 & - & - & 10.573 & - & 7.0 & -\\
3.013201 & 0.000027 & 13.183 & 0.059 & - & - & 10.693 & - & 7.0 & -\\
3.016844 & 0.000029 & 13.304 & 0.057 & - & - & 10.804 & - & 7.0 & -\\
3.022814 & 0.000003 & 13.88   &  0.03              & -   & -  & 10.98     & -                    & 7.0  & -       \\
3.024223 & 0.000008 & 13.57   &  0.03              & -  & -    & 11.02     & -                    & 7.0  & -       \\
3.024813 & 0.000003 & 14.388   & 0.005               & 3.024783  & 0.000004   & 11.76     & 0.04                 & 4.9  & 0.7        \\
\end{longtable}

\begin{longtable}{lllllll}
\caption{The table reports the metal transitions we identified (other than \CIV) with the corresponding Voigt parameters. The transitions include: \NaI\ ($\lambda \lambda \ 5891, 5897$), \CaII\ ($\lambda \lambda \ 3934, 3969$), \MgII\ ($\lambda \lambda \ 2796, 2803$), \MgI\ ($\lambda$ 2852), \CrII\ ($\lambda 2056, \lambda 2062, \lambda 2066$), \ZnII\ ($\lambda 2026, \lambda 2062$), \AlIII\ ($\lambda \lambda \ 1854, 1862$), \AlII\ ($\lambda 1670$),  \SiIV\ ($\lambda \lambda \ 1393, 1402$).} \\ 
\hline \hline
 & $z$         & $\sigma_z$ & $\log{N}$ & $\sigma_{\log{N}}$ & $b$    & $\sigma_b$ \\
   &       &           &           &            &    \kms & \kms    \\ \hline \\
\endfirsthead
\caption{continued.}\\
\hline \hline
 & $z$         & $\sigma_z$ & $\log{N}$ & $\sigma_{\log{N}}$ & $b$    & $\sigma_b$ \\
   &       &           &           &            &    \kms & \kms    \\ \hline
\endhead
\\
\hline
\endfoot
NaI                   & 0.0000277 & 0.0000003 & 11.788 & 0.009  & 4.3  & 0.2   \\
NaI                   & 0.0000702 & 0.0000004 & 11.481 & 0.038  & 1.8  & 0.4   \\
NaI                   & 0.0002304 & 0.0000012 & 10.996 & 0.034  & 2.9  & 0.8   \\
CaII                  & 0.3903456 & 0.0000066 & 10.62 & 0.11  & 8.1  & 2.4   \\
CaII                  & 0.3905171 & 0.0000013 & 10.51 & 0.13  & 1.1  & 0.4   \\
CaII                  & 0.390602 & 0.000027 & 11.135 & 0.077  & 38.5 & 7.3   \\
CaII                  & 0.3907223 & 0.0000048 & 10.50 & 0.14  & 4.6  & 2.3   \\
CaII                  & 0.3909760 & 0.0000004 & 11.871 & 0.005  & 6.5  & 0.1   \\
MgII                  & 1.0586605 & 0.0000016 & 11.203 & 0.025  & 3.2  & 0.5   \\
MgII                  & 1.059090 & 0.000016 & 11.77 & 0.20  & 8.5  & 4.4   \\
MgI (2852)            & 1.0591966 & 0.0000021 & 10.267 & 0.028  & 5.5  & 0.7   \\
MgII                  & 1.0592278 & 0.0000031 & 12.556 & 0.039  & 9.6  & 1.2   \\
MgII                  & 1.059349 & 0.000012 & 11.51 & 0.25  & 4.7  & 3.9   \\
MgII                  & 1.0594583 & 0.0000048 & 12.560 & 0.086  & 5.4  & 0.9   \\
MgI (2852)            & 1.0595180 & 0.0000095 & 11.08 & 0.12  & 13.6 & 4.6   \\
MgII                  & 1.0595351 & 0.0000026 & 13.180 & 0.023  & 7.9  & 0.3   \\
MgI (2852)            & 1.059694 & 0.000080 & 10.26 & 3.22  & 1.1  & 11.1  \\
MgII                  & 1.0596956 & 0.0000009 & 12.762 & 0.014  & 5.1  & 0.2   \\
MgI (2852)            & 1.0598196 & 0.0000018 & 11.149 & 0.016  & 8.5  & 0.5   \\
MgII                  & 1.0598329 & 0.0000006 & 13.523 & 0.034  & 7.5  & 0.2   \\
MgII                  & 1.0628774 & 0.0000011 & 11.455 & 0.014  & 4.2  & 0.3   \\
MgII                  & 1.0633291 & 0.0000045 & 10.73 & 0.58  & 1.0  & 1.6   \\
MgII                  & 1.1044476 & 0.0000015 & 11.651 & 0.016  & 5.8  & 0.3   \\
MgII                  & 1.1055630 & 0.0000038 & 11.507 & 0.031  & 7.7  & 0.8   \\
AlIII                 & 1.9172378 & 0.0000022 & 11.750 & 0.017  & 6.1  & 0.4   \\
AlIII                 & 1.9177216 & 0.0000033 & 11.351 & 0.043  & 3.2  & 0.8   \\
CrII (2062)           & 1.917722 & 0.000023 & 11.35 & 0.27  & 2.2  & 7.5   \\
ZnII (2026)           & 1.917728 & 0.000017 & 10.75 & 0.23  & 1.9  & 4.5   \\
AlIII                 & 1.9179302 & 0.0000024 & 12.141 & 0.025  & 7.7  & 0.3   \\
CrII (2056,2062,2066) & 1.917946 & 0.000014 & 11.88 & 0.10  & 7.9  & 2.8   \\
ZnII (2026,2062)      & 1.9179587 & 0.0000060 & 11.406 & 0.072  & 6.1  & 1.2   \\
AlIII                 & 1.9181308 & 0.0000043 & 11.84 & 0.18  & 6.8  & 1.1   \\
ZnII (2026,2062)      & 1.9181818 & 0.0000085 & 11.16 & 0.11  & 5.0  & 1.8   \\
CrII (2056,2062,2066) & 1.918317 & 0.000023 & 11.968 & 0.095  & 14.8 & 4.3   \\
AlIII                 & 1.9183306 & 0.0000059 & 12.65 & 0.27  & 11.9 & 1.9   \\
ZnII (2026,2062)      & 1.9183621 & 0.0000065 & 11.27 & 0.22  & 0.9  & 1.1   \\
ZnII (2026,2062)      & 1.918489 & 0.000041 & 12.265 & 0.032  & 69.8 & 4.7   \\
AlIII                 & 1.918581 & 0.000027 & 12.41 & 0.59  & 13.4 & 4.2   \\
ZnII (2026,2062)      & 1.918583 & 0.000023 & 10.81 & 0.70  & 0.8  & 1.9   \\
CrII (2056,2062,2066) & 1.9185890 & 0.0000042 & 11.867 & 0.052  & 2.1  & 1.1   \\
AlIII                 & 1.91868 & 0.00043 & 12.54 & 1.33  & 32.4 & 39.4  \\
AlIII                 & 1.919037 & 0.00080 & 12.32 & 0.98  & 55.5 & 38.2  \\
AlIII                 & 1.9191303 & 0.0000031 & 11.621 & 0.068  & 5.4  & 0.8   \\
AlIII                 & 1.920163 & 0.000079 & 10.21 & 0.68  & 6.0  & 13.9  \\
AlIII                 & 2.2213275 & 0.0000061 & 11.331 & 0.051  & 4.8  & 1.1   \\
AlIII                 & 2.2216236 & 0.0000080 & 11.142 & 0.073  & 3.7  & 1.5   \\
AlII (1670)           & 2.32918 & 0.00013 & 10.97 & 2.69  & 10.9 & 23.1  \\
AlIII                 & 2.3291879 & 0.0000049 & 11.472 & 0.026  & 8.3  & 0.7   \\
AlII (1670)           & 2.329442 & 0.000023 & 11.22 & 0.92  & 6.3  & 5.9   \\
AlIII                 & 2.3294465 & 0.0000021 & 11.821 & 0.015  & 6.9  & 0.4   \\
AlII (1670)           & 2.3306350 & 0.0000023 & 11.522 & 0.014  & 8.9  & 0.4   \\
AlIII                 & 2.3306525 & 0.0000026 & 11.687 & 0.011  & 10.1 & 0.4   \\
SiII (1526)           & 2.827909 & 0.000054 & 11.41 & 0.28  & 6.7  & 6.9   \\
SiIV                  & 2.916881 & 0.000032 & 12.601 & 0.085  & 15.5 & 2.0   \\
SiIV                  & 2.9171088 & 0.0000049 & 12.50 & 0.10  & 8.0  & 0.8   \\
SiIV (1402)           & 2.860880 & 0.000012 & 12.081 & 0.034  & 14.1 & 1.4   \\
SiII (1526)           & 2.8282811 & 0.0000034 & 11.904 & 0.024  & 4.3  & 0.5   \\
SiII (1526)           & 2.8288906 & 0.0000030 & 11.95 & 0.30  & 17.3 & 1.0  
\label{tab_metals}
\end{longtable}

\section{Additional figures}
\label{sect:6_appendix}

\begin{figure*}[h!]
\centering
\includegraphics[scale=0.50]{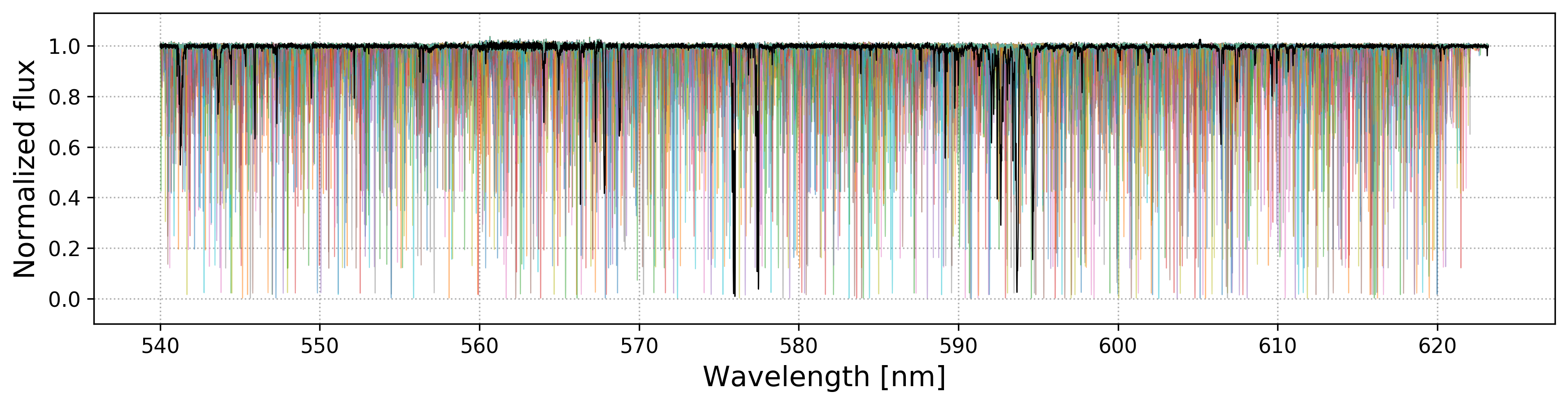}
\caption{The UVES deep spectrum (original version, no deabsorption), normalized to the continuum. Coloured lines show an example of 50 mock spectra taken from the related set of 1000 mock spectra used to estimate the uncertainties on the TPCF.}
\label{app_img:spectrum_original}
\end{figure*}

\begin{figure*}[h!]
\centering
\includegraphics[scale=0.50]{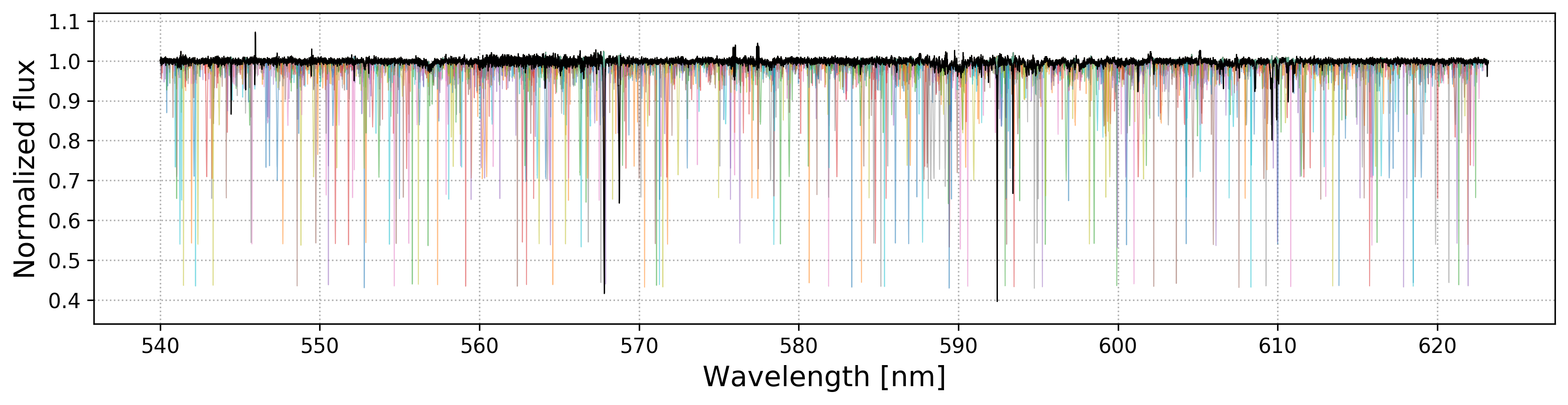}
\caption{The version of the UVES deep spectrum in which we deabsorbed all metal absorption lines and \CIV\  doublets associated with $\log N_{\rm{HI}}> 14.8$, in black, with coloured lines showing an example of 50 out of the 1000 mock spectra created for this version of the spectrum.}
\label{app_img:spectrum_14p8}
\end{figure*}

\begin{figure*}[h!]
\centering
\includegraphics[scale=0.50]{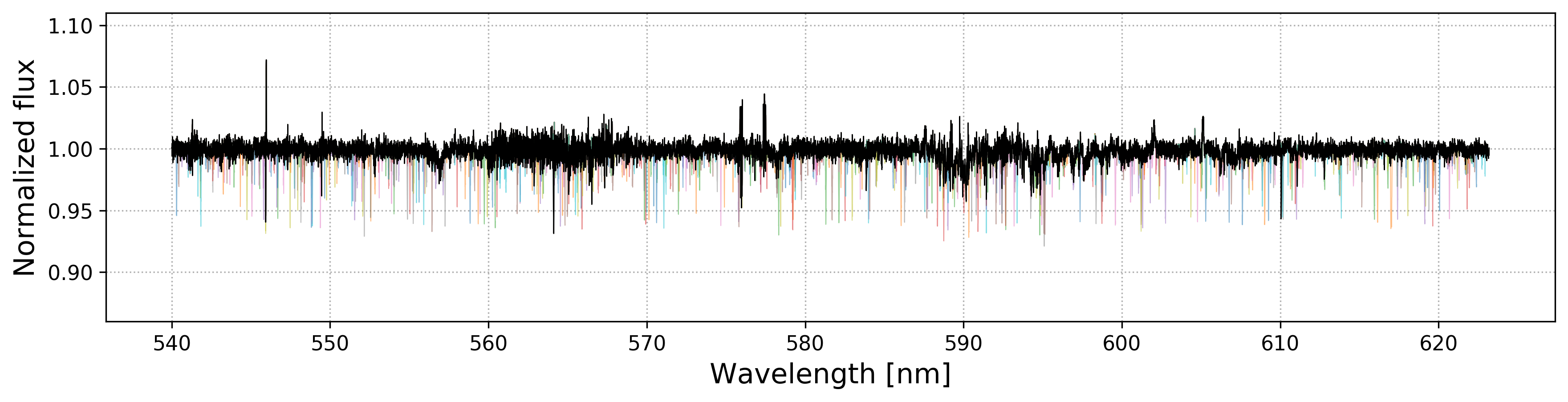}
\caption{Similar to Fig. \ref{app_img:spectrum_14p8} and Fig. \ref{app_img:spectrum_original}, but showing the case in which we left only \CIV\  associated with $\log N_{\rm{HI}}< 14.0$, in black, with an example of 50 mock spectra taken from the related set of mock spectra (coloured lines).}
\label{app_img:spectrum_14p0}
\end{figure*}

\begin{figure*}[h!]
\centering
\includegraphics[scale=0.45]{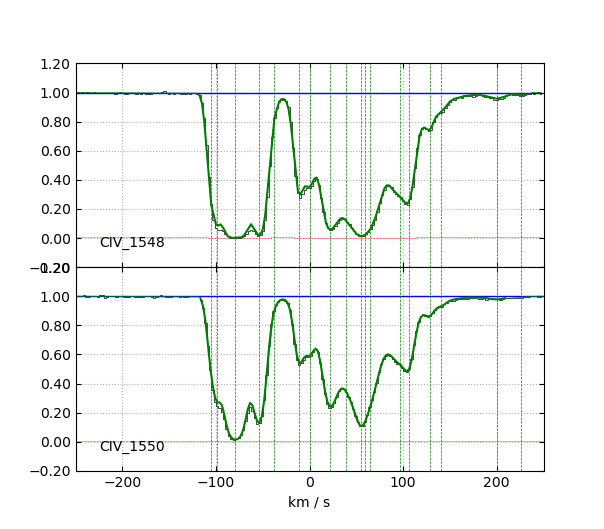}
\includegraphics[scale=0.45]{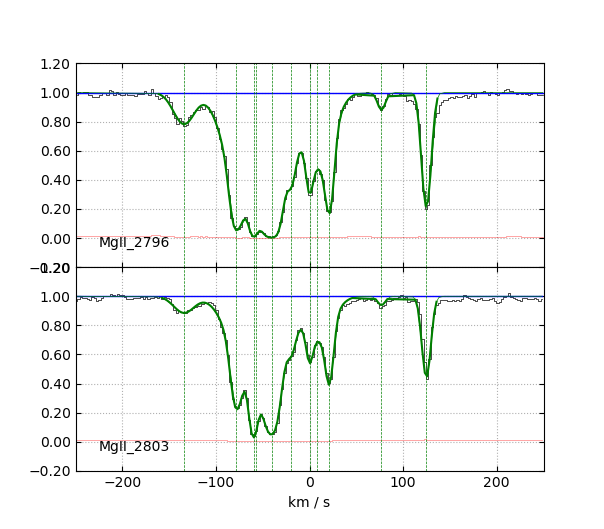}
\caption{An example of a \CIV\ system and a \MgII\ system found in the complete spectrum and fitted with the use of {\tt Astrocook}. The green line indicates the line models (Voigt fitting), the vertical dashed lines are used to mark the different components.}
\label{app_img:syst_CIV}
\end{figure*}

\end{appendix}

\end{document}